\documentclass[a4paper,12pt]{article}

\usepackage{graphicx}

\usepackage{xcolor}

\usepackage{amsmath}
\usepackage{amssymb}
\usepackage{amscd}

\usepackage{tikz-cd}

\usepackage{enumerate}
\usepackage{newtxtext,newtxmath}
\usepackage{extarrows}
\usepackage{setspace}
\usepackage{comment}
\usepackage{ulem}
\usepackage{subfigure} 
\usepackage{setspace}
\usepackage{cancel}
\usepackage{xcolor} 

\numberwithin{equation}{section}

\newcommand{\tr}{\mathrm{tr\,}}
\newcommand{\ie}{\textit{i.e., }}
\newcommand{\eg}{\textit{e.g., }}
\newcommand{\etc}{\textit{etc.}}

\begin{document}
\begin{titlepage}
\null

\vskip 1.8cm
\begin{center}

\Large \textbf{Symmetric  calorons of higher charges}\\ 
\textbf{and their large period limits}

\vskip 1.8cm
\normalsize

  {\bf Takumi Kato
  \footnote{tkato(at)st.kitasato-u.ac.jp}${}^{,a}$, Atsushi Nakamula\footnote{nakamula(at)sci.kitasato-u.ac.jp}${}^{,a}$,
 and Koki Takesue\footnote{ktakesue(at)sci.kitasato-u.ac.jp}}${}^{,b}$

\vskip 0.5cm

  { \it
  Department of Physics, School of Science \\
  Kitasato University \\
  Sagamihara, Kanagawa, 252-0373, Japan ${}^a$
  }

\vskip 0.5cm

  \textit{TECNOS Data Science Engineering Inc.\\
   Shinjuku-ku, Tokyo, 163-1427, Japan ${}^b$
  }

\vskip 2cm

\begin{abstract}
Periodic instantons, also called calorons, are the BPS solutions to the pure Yang-Mills theories on $\mathbb{R}^3\times S^1$.
It is known that the calorons interconnect with the instantons and the BPS monopoles as the ratio of their size to the period of $S^1$  varies.
We give, in this paper, the action density configurations of the $SU(2)$ calorons of higher instanton charges with several platonic symmetries through the numerical Nahm transform, after the construction of the analytic Nahm data.
The calorons considered are 5-caloron with octahedral symmetry, 7-caloron with icosahedral symmetry, 
and 4-caloron interconnecting tetrahedral and octahedral symmetries.
We also consider the large period, or the instanton, limits of the Nahm data, \ie the ADHM limits, and observe
the similar spatial distributions of the action densities with the calorons.
\end{abstract}
\end{center}
\end{titlepage}

\newpage
\section{Introduction}

Topological solitons have been attracted great attention from wide range of physics and mathematics, \eg 
\cite{MantonSutcliffe, ShifmanYung2009, Dunajski, Weinberg2012, Shnir2018}.
In four dimensional Yang-Mills gauge theories, the topological objects are instantons, which enjoy
the anti-selfdual (ASD) Yang-Mills equations, \ie 
 satulating the Bogomolnyi, or, Bogomolnyi-Plasad-Sommerfield (BPS) bound of the actions.
Particularly, on partially compactified space $\mathbb{R}^3\times S^1$,  the Yang-Mills instantons are called calorons, which can be interpreted as instantons in finite temperature \cite{HS}.
If we let the circumference of $S^1$ tend to infinity, \ie lowering the temperature, then the space 
$\mathbb{R}^3\times S^1$ turns out to be $\mathbb{R}^4$.
Hence the calorons become the instantons on this large period, or ``zero temperature" limit.

The BPS monopoles are the minimal energy solutions of three dimensional Yang-Mills-Higgs theories \cite{PS, Bogo, Shnir2005}, 
governed by the so called Bogomolnyi equations.
The equations can be derived from the dimensional reduction of four dimensional
 ASD Yang-Mills equations by imposing translational invariance
to, say, $x^0$-direction in addition to identifying the Higgs field with $A_0$.
The BPS monopoles also have a connection with the calorons.
Namely, if we identify the $S^1$ coordinate of calorons as $x^0$, and  consider the large instanton size limit with respect to the circumference 
of $S^1$, then we find the solutions to the ASD Yang-Mills equations with translational invariance in $x^0$ direction, which are 
the BPS monopoles exactly.
Here the large instanton size limit is effectively equivalent to the small size limit of $S^1$, \ie the high temperature limit.
Thus, we find that the calorons interpolate between the instantons and the BPS monopoles.

Another important class of topological solitons is Skyrmions \cite{Skyrme}.
The objects are finite energy solutions of Skyrme model, a nonlinear field theory model of pions which
successfully describes various hadrons when one introduces a suitable quantization prescription to 
the classical solutions.
The model can be interpreted as, therefore, 
a low energy effective theory of Yang-Mills gauge theories, or quantum chromo dynamics (QCD).
The Skyrme model is not a class of BPS theory, \ie the minimal energy solutions do not saturate the 
topological bound given by the second Chern number of $SU(2)$.
However, it is expected from the perspective of the nuclear physics that the Skyrmions are deeply connected to the instantons of Yang-Mills theories, which of course are BPS solutions.
In fact, Atiyah and Manton have suggested \cite{AtiyahManton} that the Skyrmions are approximately constructed from the holonomy  of the Yang-Mills instantons along a direction, say, $x^0$, with remarkable accuracy.
The explanation why the construction works successfully has been given by Sutcliffe \cite{Sutcliffe2010, Sutcliffe2015}, inspired by Sakai-Sugimoto model of holographic QCD \cite{SakaiSugimoto}.
In Sutcliffe's construction, the Yang-Mills action with a gauge potential expanded in terms of Hermite function
in $x^0$-direction gives the Skyrme energy,
 if we truncate all of the higher modes except for the lowest mode.
When the higher modes are restored, \ie introducing ``vector meson tower", the Skyrme energy
is refined and tends to BPS theory.
This Atiyah-Manton-Sutcliffe construction of the BPS Skyrme model from the Yang-Mills action is
 recently generalized to 
the case of periodic background space $\mathbb{R}^3\times S^1$, in which calorons  replacing  instantons \cite{Cork2018}.
In the latter construction, the BPS Skyrme model is naturally generalized to a family of gauged Skyrme models,
and the gauged Skyrmions are approximated by calorons, together with their monopole limits.

As mentioned, there are close relationship between instantons, calorons, BPS monopoles, and also Skyrmions,
and  the most primordial theories to those class of solitons are the Yang-Mills gauge theories, which are BPS. 
All of the solitonic objects in BPS theories are characterized by their specific topological charges,
which avoid the decay of solitons into ``fundamental particles" of the theories.
Although the Skyrmions are not BPS objects, they can be considered as approximately topological objects, 
inherited from the fundamental BPS theories.

There have been known the Skyrmions with several higher values of baryon number, which have
discrete symmetries in their spatial configurations \cite{LeeseManton1994, BattyeSutcliffe1997, BattyeSutcliffe2001, BattyeSutcliffe2002,BraatenTownsentCarson1990, FeistLauManton2013}.
A convenient method for constructing Skyrmions with platonic symmetries via a simple rational map ansatz
has been proposed in \cite{HoughtonMantonSutcliffe1998}, and it is suggested that the method has
an indirect relationship with the construction of the symmetric BPS monopoles. 
Although the higher baryon number Skyrmions are relatively well-known, a little is known about their BPS counterpart, \ie instantons \cite{SingerSutcliffe1999, Sutcliffe2004, AllenSutcliffe2013}, calorons \cite{Ward2004}, and also BPS monopoles \cite{HMM1995, HoughtonSutcliffe1996CMP, HoughtonSutcliffe1996Nonl}.
One of the motivation of this paper is to fill the gap between them. 
For this purpose, we construct $SU(2)$ calorons  with specific polyhedral symmetries of topological 
charge $N=5$ and $7$, and also $N=4$ with a moduli parameter, through the Nahm construction.
Here, the topological charge of the calorons, $N$, is identified with the corresponding 
instanton number. 
In this respect, the Nahm data of $SU(2)$ calorons of charge $N=3$ with tetrahedral symmetry
 and charge $N=4$ with octahedral symmetry are already constructed from rearranging the Nahm
 data of symmetric BPS monopoles by Ward \cite{Ward2004}.
The approach of the  present paper is to accomplish the  Ward's work to construct calorons of 
higher charges from the Nahm data of symmetric BPS monopoles.
In addition, we perform the  numerical Nahm transform of those caloron Nahm data, and also their
instanton limits, to visualize the spatial distributions of the action densities.
As mentioned, relatively small number of instantons, BPS monopoles and calorons with higher charges 
are previously known with respect to the Skyrmions, so we expect the explicit construction of 
those Yang-Mills solitons will be helpful to  the  deep understanding to those class of topological solitons.
The interrelation between them are shown in Figure 1. 

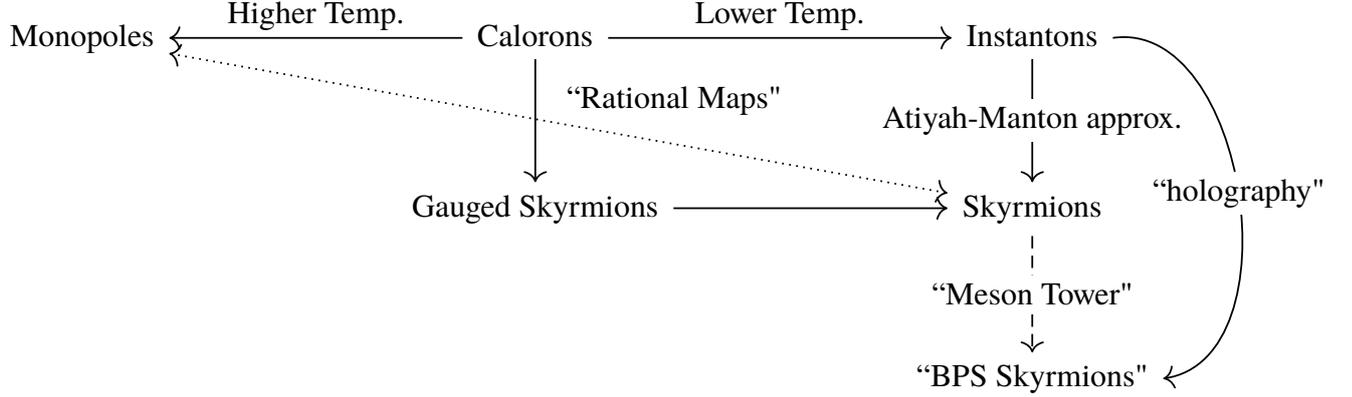
\begin{figure}[h]
\[
\begin{tikzcd}
 \mbox{Monopoles}
 \arrow[rrd,leftrightarrow,dotted, "\mbox{``Rational Maps"}"] &[20mm] 
\mbox{Calorons}
 \arrow[r, "\mbox{Lower Temp.}"]
 \arrow[l, "\mbox{Higher Temp.}"']
 \arrow[d]
 &[20mm]    
\mbox{Instantons}
 \arrow[d, "\mbox{Atiyah-Manton approx.}" description]
 \arrow[dd,bend left=90, "\mbox{``holography"}" description]\\[.8cm]
&\mbox{Gauged Skyrmions} \arrow[r]
&\mbox{Skyrmions}
\arrow[d,dashrightarrow,"\mbox{``Meson Tower"}" description]\\[.8cm]
&& \mbox{``BPS Skyrmions"}
\end{tikzcd}
\]
\caption{The interrelation between the BPS solitons to the Yang-Mills theory and the solutions to the Skyrme models.
}
\end{figure}

This paper is organized as follows.
In section 2, we give a short review on the Nahm construction of calorons and BPS monopoles with particular symmetries.
Section 3 is the main part of this paper.
In subsection 3.1, we summarize the present works on the symmetric calorons.
In subsections 3.2 through 3.4, we perform the Nahm construction of 5-caloron with octahedral symmetry, 
7-caloron with icosahedral symmetry, and 4-caloron interconnecting octahedral and tetrahedral symmetries,
respectively.
In these subsectons, the spatial configurations of each caloron are visualized by the numerical Nahm 
transform, and those of the large period, \ie instanton, limits of them are also given.
In the last subsection of section 3, we show the non-existence of 6-caloron with icosahedral symmetry.
Finally in section 4, we give concluding remarks and outlook.

\section{Symmetric Monopoles and Calorons}

In this section, we give a  brief overview on the Nahm construction for the BPS monopoles and calorons.

In the Nahm construction of the BPS monopoles \cite{Nahm1980, Hitchin1983, Nahm1984}, the fundamental ingredient is the dual ``gauge fields" to the monopole gauge fields,  the Nahm matrices.
They are triple of $N\times N$ matrix valued functions $T_j(s)\;(j=1,2,3)$ defined on a one-dimensional dual space to the configuration space $\mathbb{R}^3$, where $s\in[-1,1]$ is the dual space coordinate.
The matrix size $N$ of the Nahm data $T_j(s)$ is corresponding to the magnetic charge $N$ of the BPS monopole.
Although the gauge group is arbitrary, we restrict ourselves to the $SU(2)$ gauge theory hereafter.
The Nahm matrices are called the Nahm data if they satisfy the following conditions, which give 
 the anti-selfduality and the appropriate boundary condition for the BPS monopole fields, 
\begin{enumerate}[(a)]
\item \textit{Bulk Nahm equations:}
 \begin{align}
\frac{dT_j}{ds}(s)-\frac{i}{2}\epsilon_{jkl}[T_k(s),T_l(s)]=0, 
 \end{align}
where $s\in\left(-1,1\right)$.
\item \textit{Boundary conditions:}
 $T_j(s)$'s are regular on the interval and have poles at the boundaries $s=\pm 1$, 
 and the residues form irreducible representations of $su(2)$. \label{Monopole Boundary}

\item \textit{Hermiticity:}
 \begin{align}
T_j^\dag(s)=T_j(s),
 \end{align}

\item \textit{Reality:}
 \begin{align}
T_j(-s)={}^tT_j(s).\label{reality for monopole}
 \end{align}
\end{enumerate}

Once the Nahm data  have been obtained, we can reconstruct the gauge fields of the BPS monopole of charge $N$ through the Nahm transform defined as follows.
First, we determine the $2N$-component normalized zero-modes to the ``Dirac, or Weyl, equations" in the dual space,
 \begin{align}
\Delta^\dag u(s):=&\left(i\frac{d}{ds}1_N\otimes 1_2+T_j(s)\otimes (-i\sigma_j)+1_N\otimes x\right)u(s)=0,\\
&\int_{-1}^1u^\dag(s)u(s)ds=1_N,
 \end{align}
where $x:=x^0 \, 1_2+x^j(-i\sigma_j)$ and $1_2$ is the unit of real quaternion.
It has been shown that $\dim_{\mathbb{H}}\ker  \Delta^\dag=1$ for the  Nahm data \cite{Hitchin1983}, defined from the conditions from (a) through (d), thus the zero-mode can be determined uniquely up to gauge degrees of freedom.
Having obtained the normalized zero-mode $u(s)$, one can derive the BPS $N$-monopole gauge fields as
 \begin{align}
\Phi(\boldsymbol{x})&=\int_{-1}^1s\;u^\dag(s)u(s)ds,\\
A_j(\boldsymbol{x})&=i\int_{-1}^1u^\dag(s)\partial_ju(s)ds,
 \end{align}
where the Higgs field is defined as 
$\Phi(\boldsymbol{x}):=A_0(\boldsymbol{x})$, enjoying appropriate boundary condition.

The spatial symmetry of  BPS monopoles is encoded naturally in the Nahm data, \ie
if the monopole has specific symmetry represented by $R\in SO(3)$, then the associated Nahm data 
satisfies
\begin{align}
\left(R_N\otimes R_2\right)^{-1}\;\left(T_j(s)\otimes\sigma_j\right)\;\left(R_N\otimes R_2\right)=
T_j(s)\otimes\sigma_j,\label{Symmetric bulk data}
\end{align}
where $R_N$ and $R_2$ are the images of $R$ in $U(N)$ and $SU(2)$, respectively.

Next, we consider the Nahm construction of calorons \cite{Nahm1980},
for which  the bulk and the boundary Nahm data are involved.
The bulk data are almost the same definition for the monopole Nahm data, 
except for its defining region.
They are defined on the interval $[-\mu_0/2,\mu_0/2]$ where $\mu_0\in(0,2)$ is proportional to the inverse of the circumference of $S^1$.
When  the $SU(2)$ calorons have  an associated non-trivial holonomy, the interval is divided into two sections on which the  bulk Nahm data are defined sectorwise.
In this paper, we focus on the calorons with trivial holonomy, so that the interval is not divided and the bulk Nahm data are simply defined on the whole interval.
The boundary data $W$ is an $N$-row vector of quaternion entry,
 which enjoys the boundary Nahm equations (ii) defined below, replacing the boundary conditions for the monopoles.
 The defining conditions to the caloron Nahm data with trivial holonomy   are  given by
the following,
\begin{enumerate}[(i)]
\item \textit{Bulk Nahm equations:}
 \begin{align}
\frac{dT_j}{ds}(s)-i[T_0(s),T_j(s)]-\frac{i}{2}\epsilon_{jkl}[T_k(s),T_l(s)]=0, \label{Bulk Nahm}
 \end{align}
where $s\in\left(-\frac{\mu_0}{2},\frac{\mu_0}{2}\right)$ with $0<\mu_0/2<1$.
Note that we can always take $T_0=0$ gauge, without loss of generaligty,
 for the trivial holonomy cases.
 Hence the bulk Nahm equations are locally identical to that of the BPS monopoles.

\item \textit{Boundary Nahm equations:}
 \begin{align}
T_j(-\mu_0/2)-T_j(\mu_0/2)=\frac{1}{2}\tr_2\sigma_j W^\dag W,\label{Boundary Nahm}
 \end{align}
 
\item \textit{Hermiticity:}
 \begin{align}
T_j^\dag(s)=T_j(s),
 \end{align}

\item \textit{Reality:}
 \begin{align}
T_j(-s)={}^tT_j(s).\label{reality for caloron}
 \end{align}
\end{enumerate}

From the Nahm data of calorons, the caloron gauge fields are constructed by
the Nahm transform through the zero-mode of the ``Dirac equation" $u(s)$ and its ``boundary component" $V$,
defined as
 \begin{align}
\left\{i\frac{d}{ds}1_N\otimes 1_2-T_j(s)\otimes (-i\sigma_j)+1_N\otimes x\right\}u(s)=-W^\dag V\delta(s-\mu_0/2),
 \end{align}
with normalization
 \begin{align}
\int_{-\mu_0/2}^{\mu_0/2}u^\dag(s)u(s)ds+V^\dag V=1_N.
 \end{align}
Then the caloron gauge fields are given by
 \begin{align}
A_\mu(x_0,\boldsymbol{x})&=i\int_{-\mu_0/2}^{\mu_0/2}u^\dag(s)\partial_\mu u(s)ds+iV^\dag\partial_\mu V.
 \end{align}

As in the cases of the BPS monopoles with particular spatial symmetries,  the symmetric calorons are  given by imposing the symmetry conditions on their  Nahm data.
The condition for the bulk data is identical to (\ref{Symmetric bulk data}), and that for the boundary data is
\begin{align}
\left(R_N\otimes R_2\right)\,W^\dag=W^\dag \hat{q}_R,\label{Symmetric boundary data}
\end{align}
where $\hat{q}_R$ is a unit quaternion \cite{Ward2004}.

\section{Symmeric Calorons of Higher Charges}

In this section, we consider the caloron Nahm data and their instanton limits of some symmetric cases,
and perform their Nahm transform to visualize the symmetry of the calorons.

\subsection{Works previously studied on symmetric calorons}
Aside from the spherically symmetric calorons \cite{HS} of instanton charge 1, 
some cases of the calorons with particular spatial  symmetries 
  are already considered by several authors  for the cases of instanton charges $N=2 , 3$ and $4$.
Here we give an overview to these examples.

The simplest case of $N=2$ calorons with axial symmetry  are the Harrington-Shepard type 2-calorons 
\cite{Ward2004,Harland}, whose analytic description  of the gauge fields is given in \cite{Kato2018}.
For the case of non-trivial holonomy, the Nahm data of the charge 2-calorons with axial symmetry are 
considered in \cite{BNvB, Harland, NakamulaSakaguchi, Cork2017}, employing  the BPS 2-monopole data as their bulk data, which are given in trigonometric functions.
The  numerical Nahm transform for its trivial holonomy limit  is  performed and visualized in \cite{MNST}.

The charge $N=3$ calorons with cyclic symmetry, $C_3$, with respect to an axis are given in \cite{NakamulaSawado}, 
for which the bulk data are given in Jacobi theta functions.
The cyclic 3-calorons have a particular feature that they do not have a BPS monopole limit,
thus the Nahm data are unrelated to that of BPS monopoles.
In \cite{NakamulaSawadoTakesue}, the numerical Nahm transform of the $C_3$-symmetric calorons are performed.

\begin{figure}[h]
\begin{center}
\includegraphics[width=50mm]{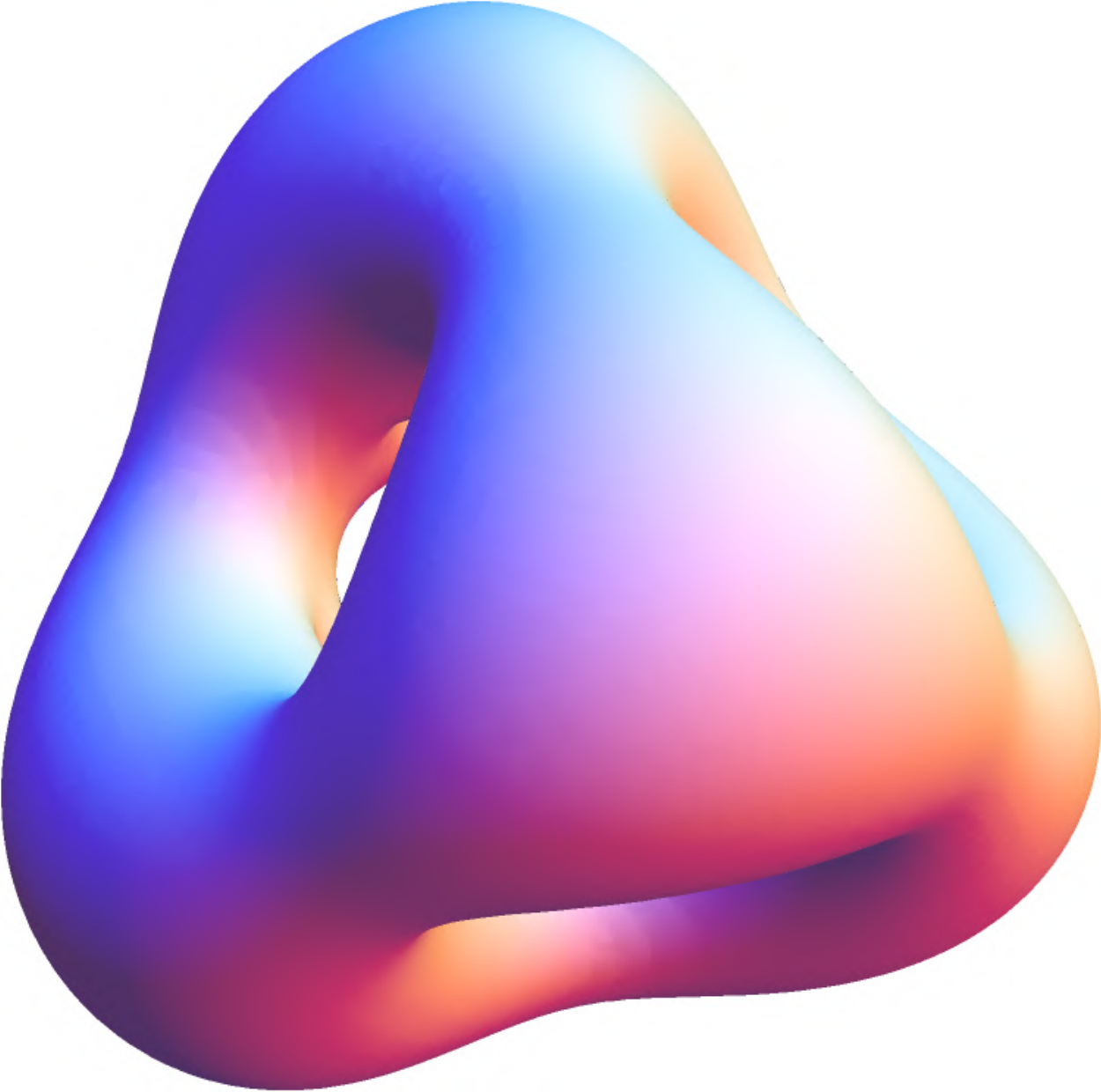}
\includegraphics[width=50mm]{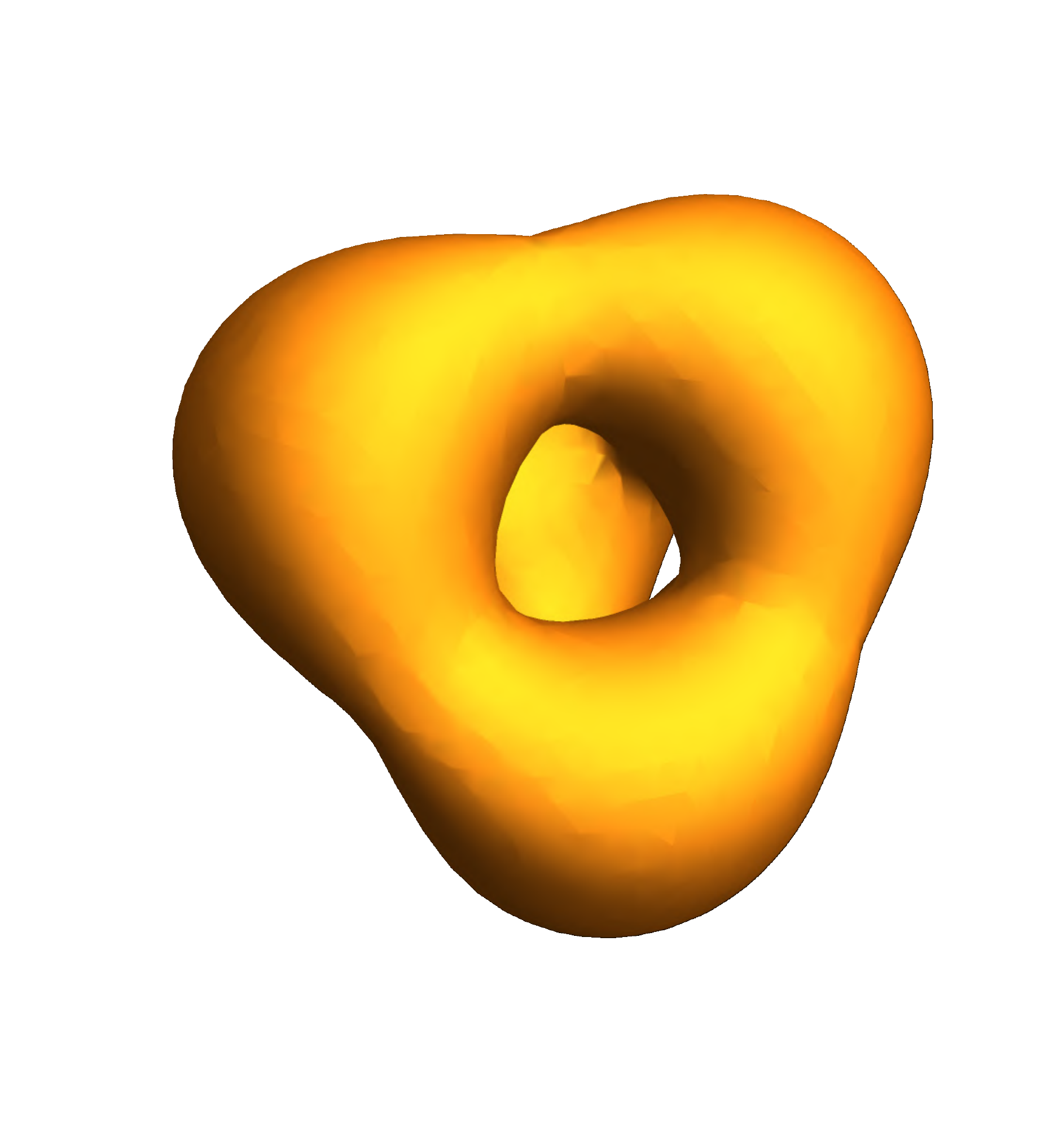}
\caption{Equi-action density surface plot for the tetrahedrally symmetric 3-caloron in $\mathbb{R}^3$ at $x_0=0$, where $\mu_0=1.0$ is applied, and the action density value is $1.3$ (left), 
and its instanton limit with $s_{\mathrm{inst}}=3.8$ (right).
The relative scale and the orientation are not significant.}\label{3-caloron}
\end{center}
\end{figure}

\begin{figure}[h]
\begin{center}
\includegraphics[width=50mm]{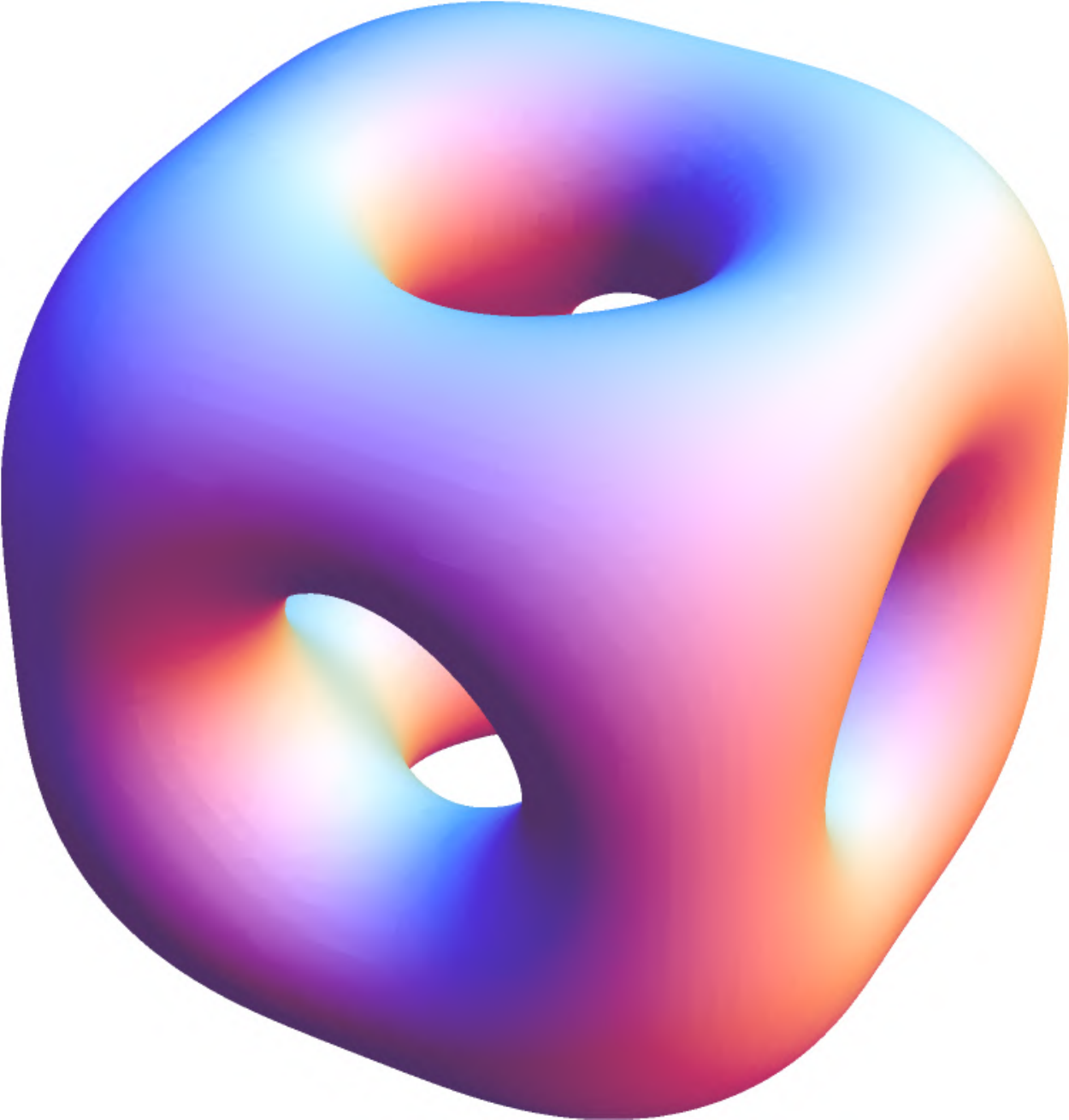}
\includegraphics[width=50mm]{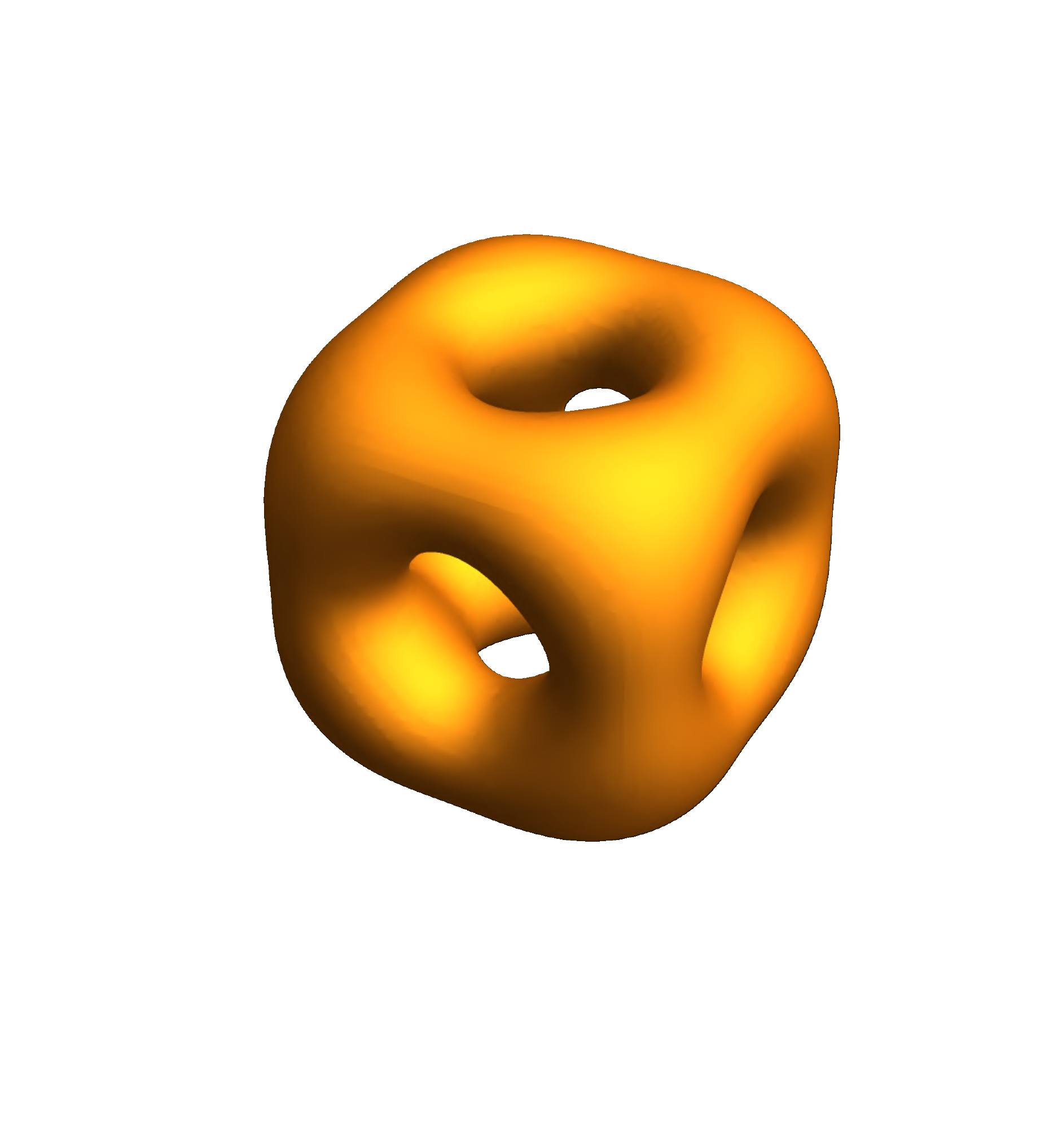}
\caption{Equi-action density surface plot for the octahedrally symmetric 4-caloron in $\mathbb{R}^3$ at $x_0=0$, where $\mu_0=1.0$ is applied, and the action density value is $0.7$ (left), 
and its instanton limit with $s_{\mathrm{inst}}=2.8$ (right).
The relative scale and the orientation are  not significant.}\label{4-caloron}
\end{center}
\end{figure}

The charge $N=3$ and $4$ calorons with tetrahedral and octahedral symmetries, respectively, are constructed by Ward \cite{Ward2004},
with the bulk data given in Weierstrass elliptic functions.
For these cases, the Nahm data of the 3-monopole with tetrahedral symmetrty, and the 4-monopole with octahedral symmetry, respectively \cite{HMM1995},  are applied to the bulk Nahm data of calorons.
Here we show the action density visualization by the numerical Nahm transform of the caloron data, 
by using the procedure  developed in \cite{MNST}.
We also give those visualization for their large period, \ie the instanton, limits in $\mathbb{R}^3$
 by using the formula 
\begin{align}
\tr F_{\mu\nu}F_{\mu\nu}\propto -\Box^2\log|\Delta^\dag\Delta|=:s_{\mathrm{inst}},\label{instanton action density}
\end{align} 
where $F_{\mu\nu}$ and $\square$ are the field strength and the Laplacian in $\mathbb{R}^4$, respectively, and $\Delta$ is the corresponding ADHM data \cite{ADHM}, see Figure \ref{3-caloron} and \ref{4-caloron}.
Both of them do not have been appeared previously.

As we have seen in this subsection, the Nahm data of calorons with specific symmetries are mostly obtained by employing the analogous BPS monopole data.
The main purpose of this paper is, therefore, devoted to confirm that the known monopole Nahm data can always be made use of the caloron Nahm data with similar symmetries.
In another word, we show that the concerning symmetric monopoles can be cooled down continuously to the lower temperature objects, \ie 
calorons and instantons.
In this respect, there are three remaining cases of the monopole Nahm data which  are not applied to the caloron Nahm data so far.
They are the octahedrally symmetric 5-monopole, the icosahedrally symmetric 7-monopole \cite{HoughtonSutcliffe1996Nonl}, 
and the one-parameter family of 4-monopoles  interconnecting tetrahedral and octahedral symmetries \cite{HoughtonSutcliffe1996CMP}, including the 4-monopole of octahedral symmetry shown in Figure \ref{4-caloron} as a special case.
In the following subsections, we construct the symmetric caloron Nahm data of charge 5, 7, and 4 with one-parameter, from the corresponding monopole data. 
We then perform the numerical Nahm transform
for the symmetric calorons to visualize their action densities. 
In addition, we investigate their large period, or instanton, limits to show that they have similar spatial symmetries of the corresponding calorons and BPS monopoles.
The final subsection  is devoted to a consideration on the non-existence of 6-caloron with icosahedral symmetry.
\newpage

\subsection{$5$-caloron with Octahedral Symmetry}
Firstly, we consider the 5-caloron with octahedral symmetry, whose bulk data are given by the Nahm data of the BPS 5-monopole with similar symmetry \cite{HoughtonSutcliffe1996Nonl}.
In this case, the bulk data enjoying (\ref{Symmetric bulk data}) are given in terms of the following matrices 
$\rho_j$ and $S_j, \;(j=1,2,3)$, which  are a generator of $so(3)$ and an octahedrally invariant vector, respectively, as
\begin{align}
\begin{array}{c}
\rho_1=
\begin{pmatrix}
0&2&0&0&0\\
2&0&\sqrt{6}&0&0\\
0&\sqrt{6}&0&\sqrt{6}&0\\
0&0&\sqrt{6}&0&2\\
0&0&0&2&0
\end{pmatrix},\quad
\rho_2=i
\begin{pmatrix}
0&-2&0&0&0\\
2&0&-\sqrt{6}&0&0\\
0&\sqrt{6}&0&-\sqrt{6}&0\\
0&0&\sqrt{6}&0&-2\\
0&0&0&2&0
\end{pmatrix},\\
\rho_3=\mathrm{diag.}\;(4,2,0,-2,-4),
\end{array}\label{rho 5}
\end{align}
and
\begin{align}
\begin{array}{c}
S_1=
\begin{pmatrix}
0&3&0&-5&0\\
3&0&-\sqrt{6}&0&-5\\
0&-\sqrt{6}&0&-\sqrt{6}&0\\
-5&0&-\sqrt{6}&0&3\\
0&-5&0&3&0
\end{pmatrix},\quad
S_2=i
\begin{pmatrix}
0&-3&0&-5&0\\
3&0&\sqrt{6}&0&-5\\
0&-\sqrt{6}&0&\sqrt{6}&0\\
5&0&-\sqrt{6}&0&-3\\
0&5&0&3&0
\end{pmatrix},\\
S_3=\mathrm{diag.}\;(-4,8,0,-8,4).
\end{array}\label{S 5}
\end{align}
These matrices enjoy the algebra
\begin{align}
[\rho_1,\rho_2]&=2i\rho_3\\
[S_1,S_2]&=-12i\rho_3-4iS_3\\
[S_1,\rho_2]+[\rho_1,S_2]&=-6iS_3,
\end{align}
together with their cyclic permutations of $1,2$ and $3$.

With (\ref{rho 5}) and (\ref{S 5}), the Nahm matries are defined as
\begin{align}
T_j(s)=x(s)\rho_j+y(s)S_j,\ (j=1,2,3),\label{bulk data N=5}
\end{align}
where $s\in(-\mu_0/2,\mu_0/2)$.
The funcitons $x(s)$ and $y(s)$ are given in terms of Weierstrass elliptic function enjoying
the differential equation
\begin{align}
\left(\wp'(u)\right)^2=4\wp^3(u)-4\wp(u),
\end{align}
as
\begin{align}
x(s)&=-\frac{\omega}{10}\left(-4\sqrt{\wp(u)}+\frac{\wp'(u)}{2\wp(u)}\right),\\
y(s)&=-\frac{\omega}{10}\left(\sqrt{\wp(u)}+\frac{\wp'(u)}{2\wp(u)}\right),
\end{align}
where  $u:=\omega(s+1)$, and  the prime denotes the derivative with respect to the argument.
The half-period, $\omega$, of the Weierstrass function in this case is
\begin{align}
\omega=\frac{1}{4\sqrt{2\pi}}\Gamma(1/4)^2.
\end{align}

The matrix basis  (\ref{rho 5}) and (\ref{S 5}), however, is not appropriate for the reality conditions (\ref{reality for monopole}) and
(\ref{reality for caloron}), thus we refer to this basis as non-reality basis. 
Although it is quite non-trivial, we can find a favorable basis for the reality conditions manifestly,
by an appropriate unitary transformation.
The reason why there exists this basis will be discussed in the final subsection.
Consequently, one of the relevant basis is given as follows,
\begin{align}
\rho_1=
\sqrt{\frac{2}{5}}
\begin{pmatrix}
0&3&0&-i&0\\
3&0&\sqrt{30}&0&-1\\
0&\sqrt{30}&0&0&0\\
i&0&0&0&3i\\
0&-1&0&-3i&0
\end{pmatrix},\
\rho_2=\sqrt{\frac{2}{5}}&
\begin{pmatrix}
0&i&0&3&0\\
-i&0&0&0&-3i\\
0&0&0&-\sqrt{30}&0\\
3&0&-\sqrt{30}&0&-1\\
0&3i&0&-1&0
\end{pmatrix},\nonumber\\
\rho_3=\frac{2}{5}
\begin{pmatrix}
-6&0&0&0&-8\\
0&0&0&5i&0\\
0&0&0&0&0\\
0&-5i&0&0&0\\
-8&0&0&0&6
\end{pmatrix}&,\label{octa5 reality basis rho}
\end{align}
and
\begin{align}
S_1=\sqrt{\frac{2}{5}}
\begin{pmatrix}
0&-3&0&-4i&0\\
-3&0&-\sqrt{30}&0&1\\
0&-\sqrt{30}&0&0&0\\
4i&0&0&0&12i\\
0&1&0&-12i&0
\end{pmatrix},\
S_2=\sqrt{\frac{2}{5}}&
\begin{pmatrix}
0&4i&0&-3&0\\
-4i&0&0&0&-12i\\
0&0&0&\sqrt{30}&0\\
-3&0&\sqrt{30}&0&1\nonumber \\
0&12i&0&1&0
\end{pmatrix},\\
S_3=\frac{4}{5}
\begin{pmatrix}
3&0&0&0&4\\
0&0&0&10i&0\\
0&0&0&0&0\\
0&-10i&0&0&0\\
4&0&0&0&-3
\end{pmatrix}&.\label{octa5 reality basis S}
\end{align}
It is straightforward to confirm that this bulk data enjoy the bulk Nahm equaitons, the hermiticity and the reality.
We refer the basis (\ref{octa5 reality basis rho}) and (\ref{octa5 reality basis S}) as the reality basis.

The next task we have to do is to find the solution to the  boundary Nahm equations (\ref{Boundary Nahm}), in the reality basis.
A straightforward calculation shows that the solution is,
\begin{align}
W=\lambda\left(\mathbf{1}_2,i\sqrt{10}\sigma_2,0,-i\sqrt{10}\sigma_1,3\cdot\mathbf{1}_2\right),
\label{Octa5 boundary data}
\end{align}
where
\begin{align}
\lambda=\sqrt{-\frac{2}{5}x(\mu_0/2)-\frac{8}{5}y(\mu_0/2)}.\label{Octa5 lambda}
\end{align}
Here $\lambda$ takes real value because $x(s)$ and $y(s)$ are negative for $s>0$.
Hence, we have found the Nahm data of the octahedrally symmetric 5-caloron explicitly. 

\begin{figure}[h]
\begin{center}
\includegraphics[width=55mm]{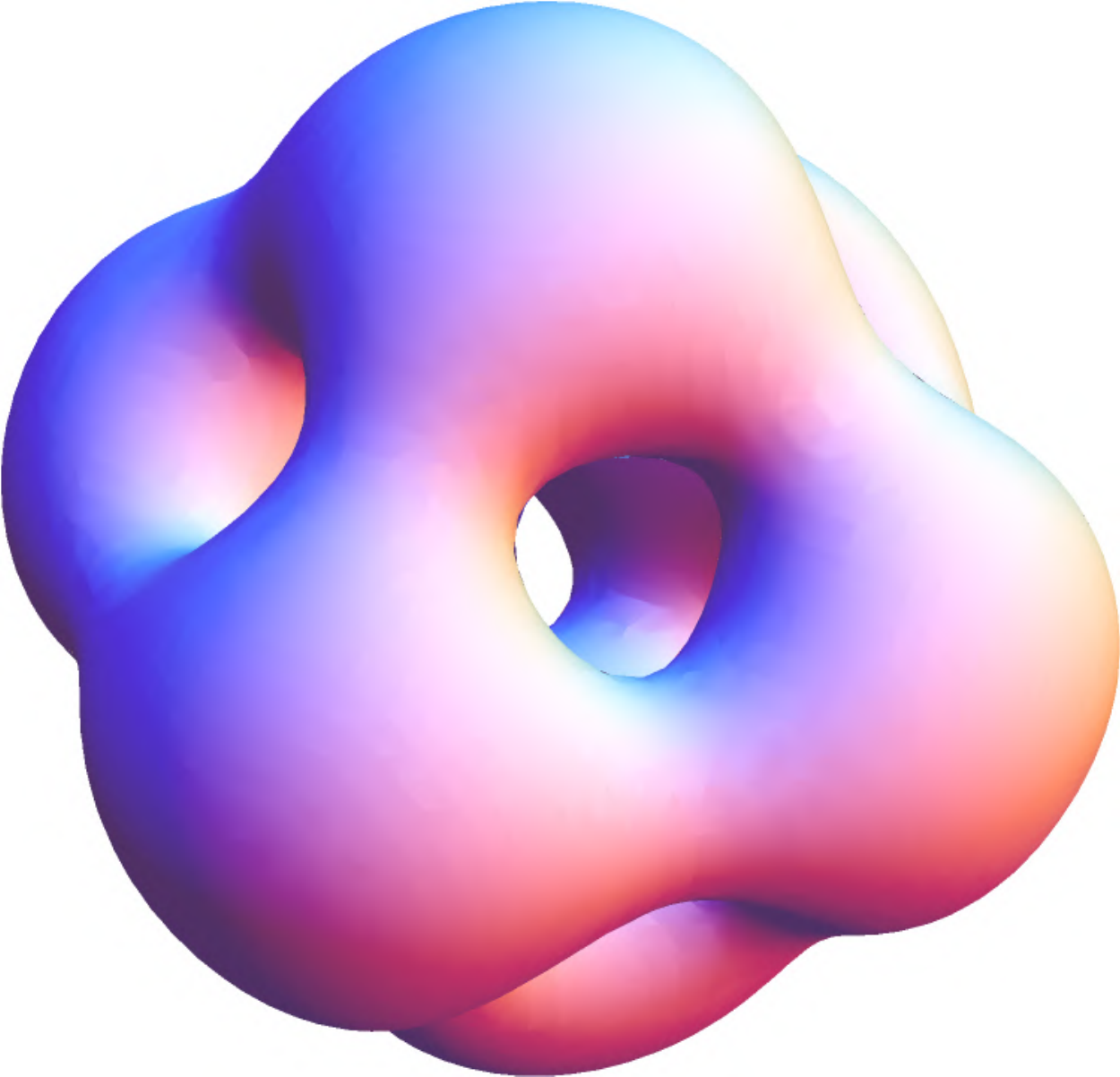}
\caption{Equi-action density surface plot for the 5-caloron in $\mathbb{R}^3$ at $x_0=0$, with $\mu_0=1.6$, and the action density value is $0.13$.}\label{Octa5 caloron}
\end{center}
\end{figure}

Performing the scheme of numerical Nahm transform \textit{\'a la} \cite{MNST}, we show the equi-action density surface plot in $\mathbb{R}^3$ at $x_0=0$, in Figure \ref{Octa5 caloron}, where $\mu_0=1.6$ is applied.

\begin{figure}[h]
\begin{center}
\includegraphics[width=70mm]{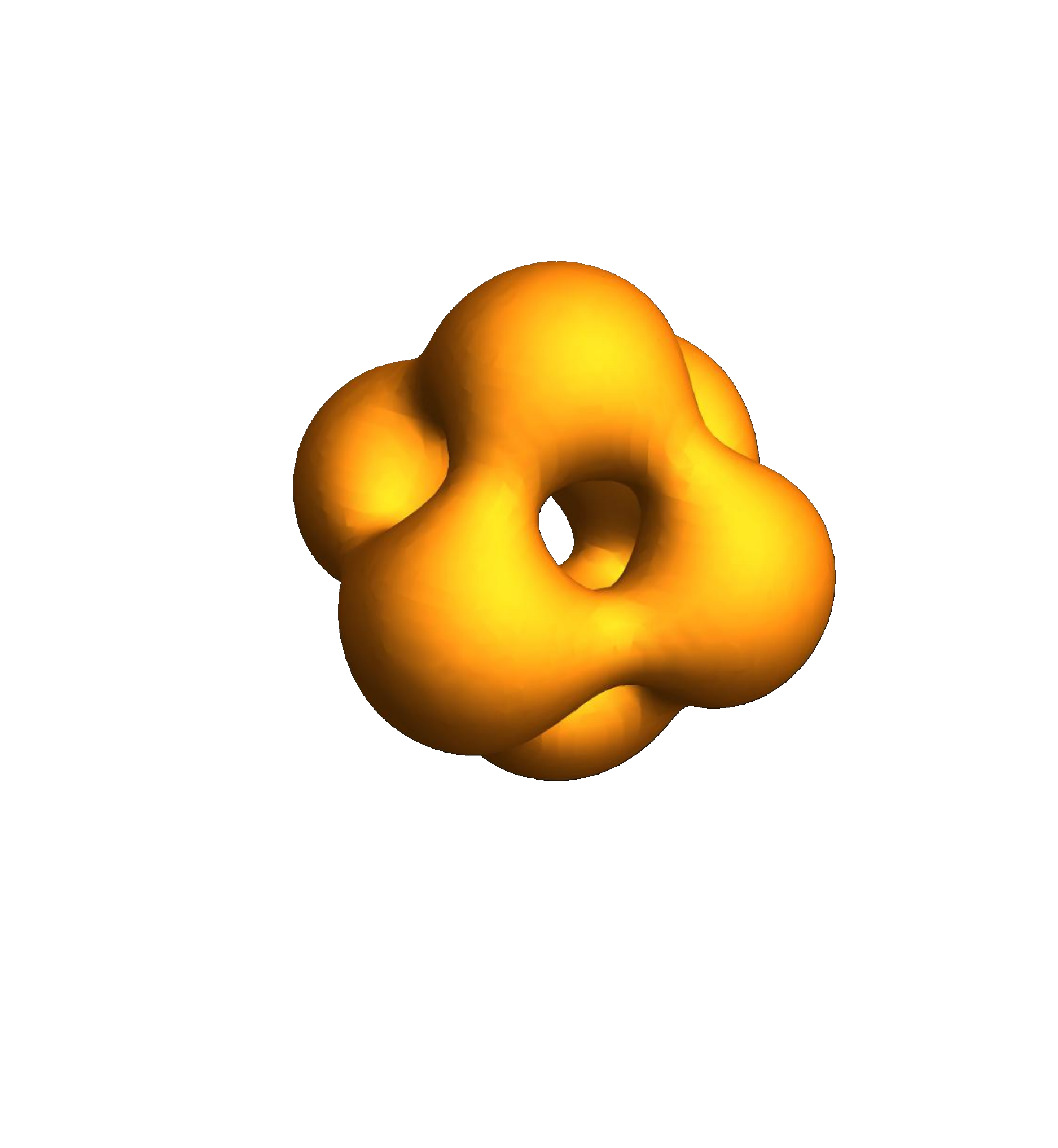}
\caption{Equi-action density surface plot for the instanton limit of the 5-caloron in $\mathbb{R}^3$ at $x_0=0$, and $s_{\mathrm{inst}}=1.06$.
The relative scale and the orientation with respect to Figure \ref{Octa5 caloron} are not significant.}\label{Octa5 instanton}
\end{center}
\end{figure}

The instanton, or the large period ($S^1\to\mathbb{R}$) limit of the Nahm data can be derived from the $\mu_0\to0$ limit of the Nahm data.
From the behavior $\lim_{\mu_0\to0}\lambda/\sqrt{\mu_0}=\omega/\sqrt{5}$ of the boundary data (\ref{Octa5 boundary data}) and (\ref{Octa5 lambda}), together with the value $T_j(0)$, we find
the corresponding ADHM matrix \cite{ADHM} is  
\begin{align}
\Delta=\frac{\omega}{\sqrt{5}}
\begin{pmatrix}
\mathbf{1}_2 &i\sqrt{10}\sigma_2&0&-i\sqrt{10}\sigma_1&3\cdot \mathbf{1}_2\\
-\frac{6}{\sqrt{5}}i\sigma_3 &\frac{3}{\sqrt{2}}i\sigma_1&0&\frac{3}{\sqrt{2}}i\sigma_2& -\frac{8}{\sqrt{5}}i\sigma_3\\
\frac{3}{\sqrt{2}}i\sigma_1 &0&\sqrt{15}i\sigma_1&0& \frac{1}{\sqrt{2}}i\sigma_1\\
0 &\sqrt{15}i\sigma_1 & 0 & -\sqrt{15}i\sigma_2 & 0\\
\frac{3}{\sqrt{2}}i\sigma_2 & 0 & -\sqrt{15}i\sigma_2 & 0 &-\frac{1}{\sqrt{2}}i\sigma_2\\
-\frac{8}{\sqrt{5}}i\sigma_3 &-\frac{1}{\sqrt{2}}i\sigma_1&0&-\frac{1}{\sqrt{2}}i\sigma_2&\frac{6}{\sqrt{5}}i\sigma_3
\end{pmatrix}+
\begin{pmatrix}
0&0&0&0&0\\
x&0&0&0&0\\
0&x&0&0&0\\
0&0&x&0&0\\
0&0&0&x&0\\
0&0&0&0&x
\end{pmatrix}.
\end{align}
Here we show the equi-action density surface plot in $\mathbb{R}^3$ at $x_0=0$, by using (\ref{instanton action density}) in Figure \ref{Octa5 instanton}.

We find the spatial symmetry of the 5-instanton is similar to the corresponding caloron as well as the BPS monopole,
and observe that the 5-monopole with octahedral symmetry can smoothly be cooled down into instanton through the 5-caloron.

\subsection{$7$-caloron with Icosahedral Symmetry}

In this subsection, we consider the icosahedrally symmetric 7-caloron.
As in  the last subsection, we apply the Nahm data of the  icosahedral 7-monopole \cite{HoughtonSutcliffe1996Nonl} as the bulk Nahm data.
The relevant 7-dimensional representations of  $\rho_j$ and $S_j$, the generators of $so(3)$ and icosahedrally invariant vector, respectively, are given as follows,
\begin{align}
\rho_1&=\begin{pmatrix}
0&\sqrt{6}&0&0&0&0&0\\
\sqrt{6}&0&\sqrt{10}&0&0&0&0\\
0&\sqrt{10}&0&2\sqrt{3}&0&0&0\\
0&0&2\sqrt{3}&0&2\sqrt{3}&0&0\\
0&0&0&2\sqrt{3}&0&\sqrt{10}&0\\
0&0&0&0&\sqrt{10}&0&\sqrt{6}\\
0&0&0&0&0&\sqrt{6}&0
\end{pmatrix},\nonumber\\
\rho_2&=i
\begin{pmatrix}
0&-\sqrt{6}&0&0&0&0&0\\
\sqrt{6}&0&-\sqrt{10}&0&0&0&0\\
0&\sqrt{10}&0&-2\sqrt{3}&0&0&0\\
0&0&2\sqrt{3}&0&-2\sqrt{3}&0&0\\
0&0&0&2\sqrt{3}&0&-\sqrt{10}&0\\
0&0&0&0&\sqrt{10}&0&-\sqrt{6}\\
0&0&0&0&0&\sqrt{6}&0
\end{pmatrix},\nonumber\\
\rho_3&=\mathrm{diag.}\;(6,4,2,0,-2,-4,-6)\label{rho 7}
\end{align}
and
\begin{align}
S_1&=
\begin{pmatrix}
0&-5\sqrt{6}&0&0&14\sqrt{15}&0&0\\
-5\sqrt{6}&0&9\sqrt{10}&0&0&0&0\\
0&9\sqrt{10}&0&-10\sqrt{3}&0&0&-14\sqrt{15}\\
0&0&-10\sqrt{3}&0&-10\sqrt{3}&0&0\\
14\sqrt{15}&0&0&-10\sqrt{3}&0&9\sqrt{10}&0\\
0&0&0&0&9\sqrt{10}&0&-5\sqrt{6}\\
0&0&-14\sqrt{15}&0&0&-5\sqrt{6}&0
\end{pmatrix},\nonumber\\
S_2&=i
\begin{pmatrix}
0&5\sqrt{6}&0&0&14\sqrt{15}&0&0\\
-5\sqrt{6}&0&-9\sqrt{10}&0&0&0&0\\
0&9\sqrt{10}&0&10\sqrt{3}&0&0&-14\sqrt{15}\\
0&0&-10\sqrt{3}&0&10\sqrt{3}&0&0\\
-14\sqrt{15}&0&0&-10\sqrt{3}&0&-9\sqrt{10}&0\\
0&0&0&0&9\sqrt{10}&0&5\sqrt{6}\\
0&0&14\sqrt{15}&0&0&-5\sqrt{6}&0
\end{pmatrix},\\
S_3&=\begin{pmatrix}
12&0&0&0&0&14\sqrt{6}&0\\
0&-48&0&0&0&0&14\sqrt{6}\\
0&0&60&0&0&0&0\\
0&0&0&0&0&0&0\\
0&0&0&0&-60&0&0\\
14\sqrt{6}&0&0&0&0&48&0\\
0&0&14\sqrt{6}&0&0&0&-12
\end{pmatrix}.\nonumber\label{S 7}
\end{align}
The algebra for these matrices  is
\begin{align}
[\rho_1,\rho_2]&=2i\rho_3,\\
[S_1,S_2]&=-750i\rho_3-90iS_3,\\
[S_1,\rho_2]+[\rho_1,S_2]&=-10iS_3,
\end{align}
and their cyclic permutations of $1,2$ and $3$.

With this matrix basis (\ref{rho 7}) and (\ref{S 7}), the bulk Nahm data is defined as 
\begin{align}
T_j(s)=x(s)\rho_j+y(s)S_j, \ (j=1,2,3)\label{bulk data N=7}
\end{align}
where the defining interval  of $s$ is identical to the previous case.
The solution to the bulk Nahm equations (\ref{Bulk Nahm}) is  given by the Weierstrass function satisfying
\begin{align} 
\left(\wp'(u)\right)^2=4\wp^3(u)-4,
\end{align}
and we find
\begin{align}
x(s)&=-\frac{\omega}{14}\left(-6\sqrt{\wp(u)}+\frac{\wp'(u)}{2\wp(u)}\right),\\
y(s)&=-\frac{\omega}{70}\left(\sqrt{\wp(u)}+\frac{\wp'(u)}{2\wp(u)}\right),
\end{align}
where $u=\omega(s+1)$ with the half-period of the Weierstrass function
\begin{align}
 \omega=\frac{1}{4\sqrt{3\pi}}\Gamma(1/3)\Gamma(1/6).
\end{align}

As in the previous case, the representation (\ref{rho 7}) and (\ref{S 7}) is not suitable for the reality conditions (\ref{reality for caloron}), \ie  a non-reality basis. 
Similarly to the 5-caloron case, we can find an appropriate basis for the reality conditions  through a unitary transformation, that is,
\begin{align}
\rho_1&=\frac{1}{5}
\begin{pmatrix}
-12&0&2i\sqrt{15}&0&2\sqrt{15}&-\sqrt{6}&0\\
0&12&3\sqrt{10}&0&-2i\sqrt{10}&0&\sqrt{6}\\
-2i\sqrt{15}&3\sqrt{10}&0&10\sqrt{6}&0&2i\sqrt{10}&-2\sqrt{15}\\
0&0&10\sqrt{6}&0&0&0&0\\
2\sqrt{15}&2i\sqrt{10}&0&0&0&3\sqrt{10}&2i\sqrt{15}\\
-\sqrt{6}&0&-2i\sqrt{10}&0&3\sqrt{10}&12&0\\
0&\sqrt{6}&-2\sqrt{15}&0&-2i\sqrt{15}&0&-12
\end{pmatrix},\nonumber\\
\rho_2&=\frac{1}{\sqrt{5}}
\begin{pmatrix}
0&\sqrt{30}&-2\sqrt{3}&0&2i\sqrt{3}&0&0\\
\sqrt{30}&0&2i\sqrt{2}&0&3\sqrt{2}&0&0\\
-2\sqrt{3}&-2i\sqrt{2}&0&0&0&-3\sqrt{2}&-2i\sqrt{3}\\
0&0&0&0&-2\sqrt{30}&0&0\\
-2i\sqrt{3}&3\sqrt{2}&0&-2\sqrt{30}&0&2i\sqrt{2}&-2\sqrt{3}\\
0&0&-3\sqrt{2}&0&-2i\sqrt{2}&0&\sqrt{30}\\
0&0&2i\sqrt{3}&0&-2\sqrt{3}&\sqrt{30}&0
\end{pmatrix},\\
\rho_3&=\frac{2}{5}
\begin{pmatrix}
-12&i\sqrt{6}&0&0&0&-\sqrt{6}&3i\\
-i\sqrt{6}&-12&0&0&0&2i&-\sqrt{6}\\
0&0&0&0&5i&0&0\\
0&0&0&0&0&0&0\\
0&0&-5i&0&0&0&0\\
-\sqrt{6}&-2i&0&0&0&12&-i\sqrt{6}\\
-3i&-\sqrt{6}&0&0&0&i\sqrt{6}&12
\end{pmatrix}\nonumber
\end{align}
and
\begin{align}
S_1&=\frac{1}{5}
\begin{pmatrix}
60&0&60i\sqrt{15}&0&-10\sqrt{15}&5\sqrt{6}&0\\
0&-60&-15\sqrt{10}&0&-60i\sqrt{10}&0&-5\sqrt{6}\\
-60i\sqrt{15}&-15\sqrt{10}&0&-50\sqrt{6}&0&60i\sqrt{10}&10\sqrt{15}\\
0&0&-50\sqrt{6}&0&0&0&0\\
-10\sqrt{15}&60i\sqrt{10}&0&0&0&-15\sqrt{10}&60i\sqrt{15}\\
5\sqrt{6}&0&-60i\sqrt{10}&0&-15\sqrt{10}&-60&0\\
0&-5\sqrt{6}&10\sqrt{15}&0&-60i\sqrt{15}&0&60
\end{pmatrix},\nonumber\\
S_2&=\frac{1}{\sqrt{5}}
\begin{pmatrix}
0&-5\sqrt{30}&10\sqrt{3}&0&60i\sqrt{3}&0&0\\
-5\sqrt{30}&0&60i\sqrt{2}&0&-15\sqrt{2}&0&0\\
10\sqrt{3}&-60i\sqrt{2}&0&0&0&15\sqrt{2}&-60i\sqrt{3}\\
0&0&0&0&10\sqrt{30}&0&0\\
-60i\sqrt{3}&-15\sqrt{2}&0&10\sqrt{30}&0&60i\sqrt{2}&10\sqrt{3}\\
0&0&15\sqrt{2}&0&-60i\sqrt{2}&0&-5\sqrt{30}\\
0&0&60i\sqrt{3}&0&10\sqrt{3}&-5\sqrt{30}&0
\end{pmatrix},\\
S_3&=\frac{2}{5}\begin{pmatrix}
60&30i\sqrt{6}&0&0&0&5\sqrt{6}&90i\\
-30i\sqrt{6}&60&0&0&0&60i&5\sqrt{6}\\
0&0&0&0&150i&0&0\\
0&0&0&0&0&0&0\\
0&0&-150i&0&0&0&0\\
5\sqrt{6}&-60i&0&0&0&-60&-30i\sqrt{6}\\
-90i&5\sqrt{6}&0&0&0&30i\sqrt{6}&-60
\end{pmatrix}.\nonumber
\end{align}

\begin{figure}[h]
\begin{center}
\includegraphics[width=70mm]{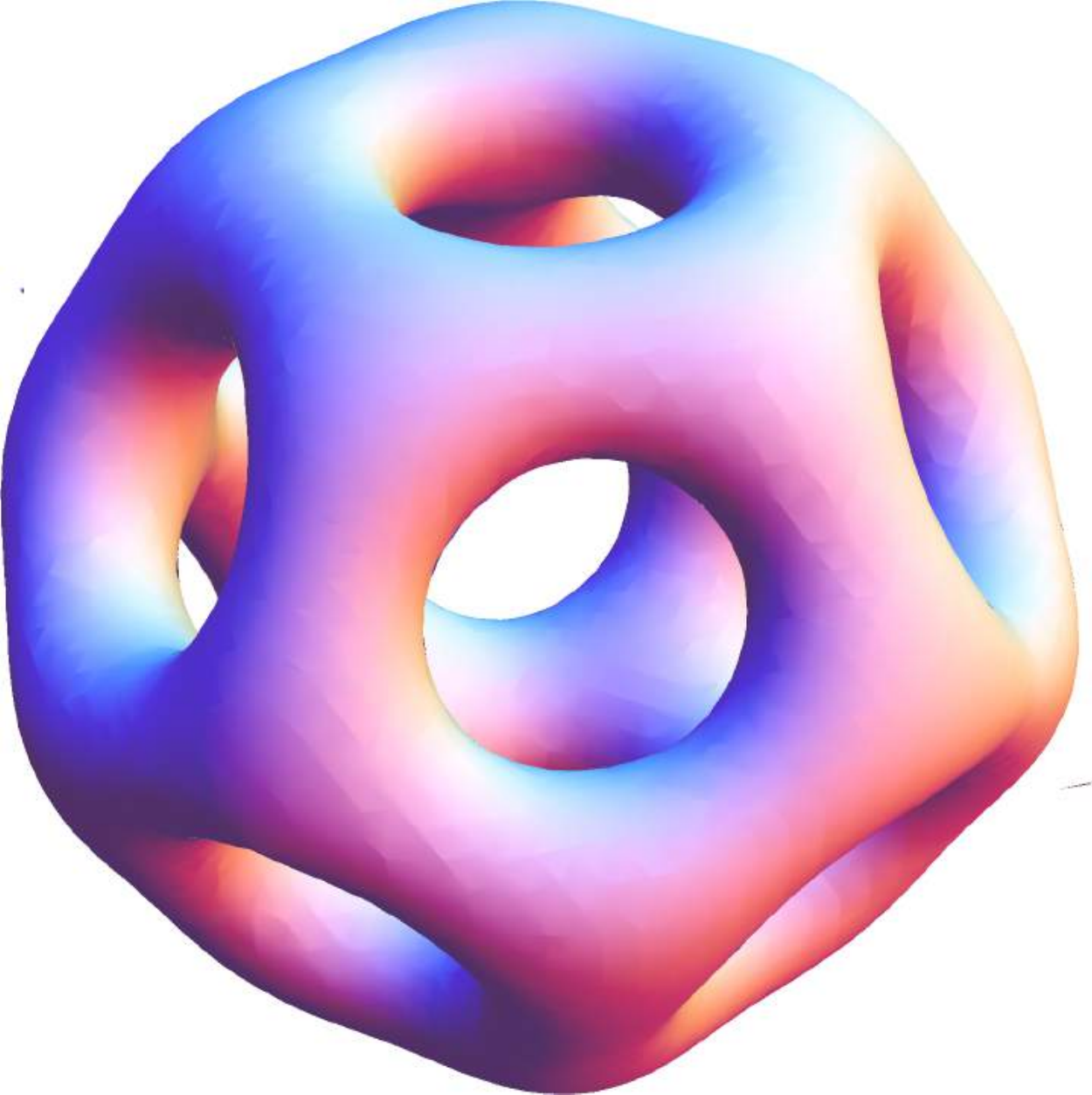}
\caption{Plot of the equi-action density surface for the 7-caloron in $\mathbb{R}^3$ at $x_0=0$, with $\mu_0=1.6$, and the action density value is $0.11$.}\label{Icosa7 caloron}
\end{center}
\end{figure}

The solution to the boundary Nahm equations (\ref{Boundary Nahm}) in this reality basis is found to be
\begin{align}
W=\lambda\left(\mathbf{1}_2,i\sqrt{\frac{2}{3}}\sigma_3,i\sqrt{\frac{5}{3}}\sigma_1,0,i\sqrt{\frac{5}{3}}\sigma_2,-\sqrt{\frac{2}{3}}\mathbf{1}_2,
i\sigma_3\right),
\end{align}
where
\begin{align}
\lambda=2\sqrt{\frac{-3}{5}\left\{x(\mu_0/2)+30y(\mu_0/2)\right\}}.
\end{align}
The shape of this 7-caloron with icosahedral symmetry can be visualized by performing the numerical Nahm transformation, as in the previous cases.
We show the equi-action density surface plot in $\mathbb{R}^3$ at $x_0=0$ in Figure \ref{Icosa7 caloron}, and find a dodecahedral structure as in the 7-monopole case, where $\mu_0=1.6$ is applied.

\begin{figure}[h]
\begin{center}
\includegraphics[width=70mm]{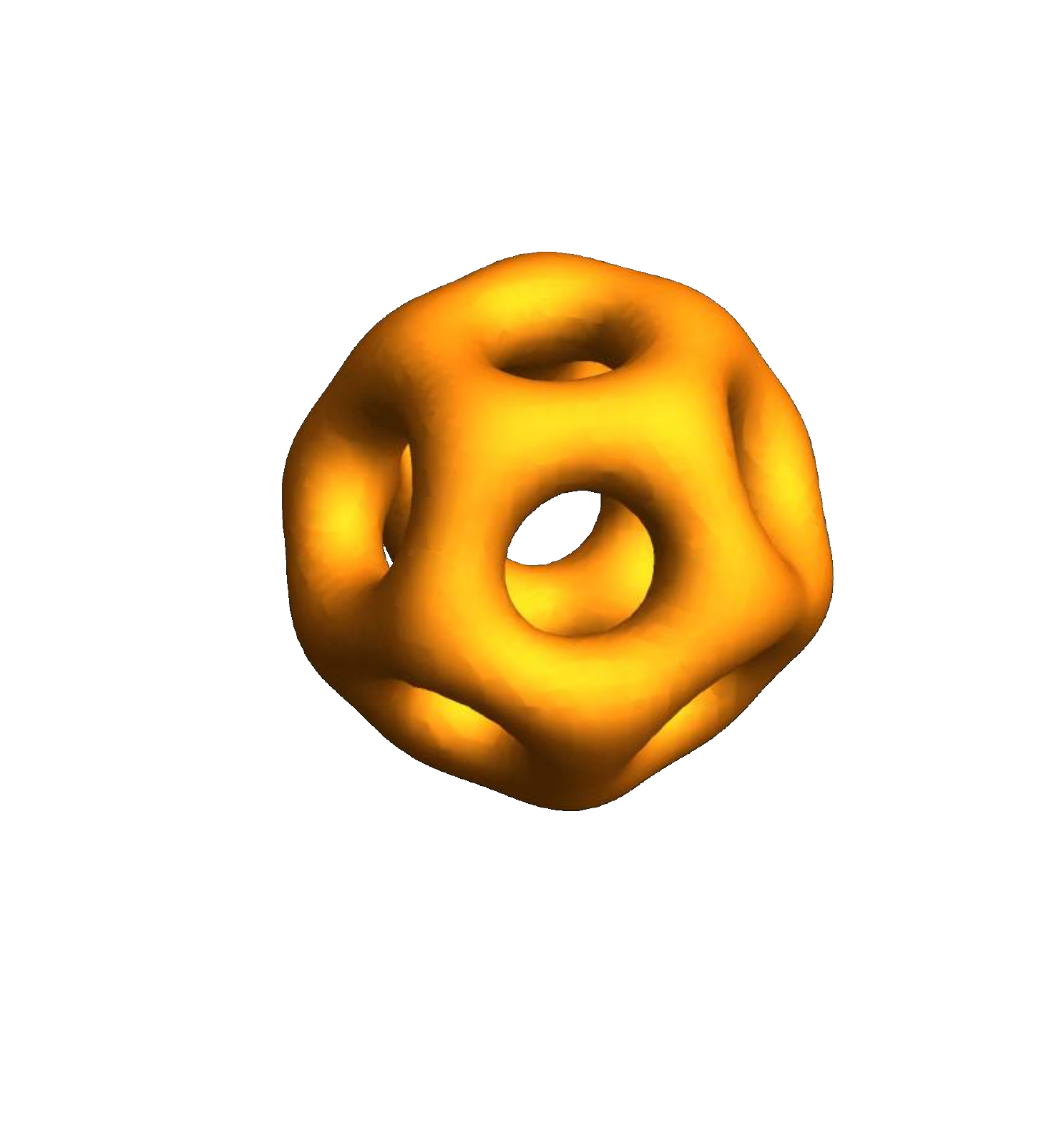}
\caption{Plot of the equi-action density surface for the instanton limit of the 7-caloron in $\mathbb{R}^3$ at $x_0=0$, and $s_{\mathrm{inst}}=1.1$.
The relative scale and the orientation with respect to Figure \ref{Icosa7 caloron} are not significant.}\label{Icosa7 instanton}
\end{center}
\end{figure}

The instanton limit of the icosahedral 7-caloron leads to the ADHM data as
\begin{align}
\Delta=\frac{3\omega}{\sqrt{5}}&
\begin{pmatrix}
\mathbf{1}_2 &i\sqrt{\frac{2}{3}}\sigma_3&i\sqrt{\frac{5}{3}}\sigma_1&0&i\sqrt{\frac{5}{3}}\sigma_2&-\sqrt{\frac{2}{3}}
\mathbf{1}_2&i\sigma_3\\
-\frac{2(i\sigma_1+2i\sigma_3)}{\sqrt{5}} &\frac{5}{\sqrt{6}}i\sigma_2&-\frac{1}{\sqrt{3}}i\sigma_2&0&\frac{1}{\sqrt{3}}i\sigma_1
& -\frac{i\sigma_1+2i\sigma_3}{\sqrt{30}} &0 \\
\sqrt{\frac{5}{6}}i\sigma_2&\frac{2(i\sigma_1-2i\sigma_3)}{\sqrt{5}} &\frac{1}{\sqrt{2}}i\sigma_1&0&\frac{1}{\sqrt{2}}i\sigma_2
&0& \frac{i\sigma_1-2i\sigma_3}{\sqrt{30}} \\
-\frac{1}{\sqrt{3}}i\sigma_2&\frac{1}{\sqrt{2}}i\sigma_1 &0&\sqrt{\frac{10}{3}}i\sigma_1&0&-\frac{1}{\sqrt{2}}i\sigma_2
& -\frac{1}{\sqrt{3}}i\sigma_1 \\
0&0&\sqrt{\frac{10}{3}}i\sigma_1&0&-\sqrt{\frac{10}{3}}i\sigma_2&0&0\\
\frac{1}{\sqrt{3}}i\sigma_1&\frac{1}{\sqrt{2}}i\sigma_2 &0&-\sqrt{\frac{10}{3}}i\sigma_2&0&\frac{1}{\sqrt{2}}i\sigma_1
& -\frac{1}{\sqrt{3}}i\sigma_2 \\
-\frac{i\sigma_1+2i\sigma_3}{\sqrt{30}} &0&-\frac{1}{\sqrt{2}}i\sigma_2&0&\frac{1}{\sqrt{2}}i\sigma_1
& \frac{2(i\sigma_1+2i\sigma_3)}{\sqrt{5}}&\sqrt{\frac{5}{6}}i\sigma_2 \\
0&\frac{i\sigma_1-2i\sigma_3}{\sqrt{30}} &-\frac{1}{\sqrt{3}}i\sigma_1&0&-\frac{1}{\sqrt{3}}i\sigma_2
&\sqrt{\frac{5}{6}}i\sigma_2& -\frac{2(i\sigma_1-2i\sigma_3)}{\sqrt{5}} \\
\end{pmatrix}\nonumber\\&+
\begin{pmatrix}
0&0&0&0&0&0&0\\
x&0&0&0&0&0&0\\
0&x&0&0&0&0&0\\
0&0&x&0&0&0&0\\
0&0&0&x&0&0&0\\
0&0&0&0&x&0&0\\
0&0&0&0&0&x&0\\
0&0&0&0&0&0&x
\end{pmatrix}.
\end{align}
We find that the action density distribution in $\mathbb{R}^3$ at $x_0=0$ also has dodecahedrally symmetric shape, by using (\ref{instanton action density}) in Figure \ref{Icosa7 instanton}.

We therefore observe that the 7-momopole with icosahedral symmetry can be cooled down smoothly into the 7-instanton through the 7-caloron.

\subsection{One parameter family of $4$-caloron} \label{One parameter family}

In this subsection, we consider a one parameter family of 4-caloron interpolating between 
tetrahedrally and octahedrally symmetric configurations, as a periodic generalization of 
the BPS 4-monopole with similar structure \cite{HoughtonSutcliffe1996CMP}.
As mentioned earlier, this gives a generalization of the octahedral 4-caloron \cite{Ward2004}.

For this family of calorons, the relevant algebra for the bulk Nahm data are spanned by $\rho_j,\;S_{T, \,j}$ and $S_{O, \,j}\;(j=1,2,3)$.
Here, as in the previous cases, $\rho_j$ is a generator of $so(3)$, and $S_{T,\,j}$ and $S_{O, \,j}$ are 
tetrahedrally and octahedrally invariant vectors, respectively.
We give, without derivation, a 4-dimensional representation in a basis appropriate for the reality condition, as follows,
\begin{align}
\begin{array}{c}
\rho_1=\begin{pmatrix}
-\sqrt{3}&0&-i&-1\\
0&\sqrt{3}&-1&i\\
i&-1&-\sqrt{3}&0\\
-1&-i&0&\sqrt{3}
\end{pmatrix},
\quad
\rho_3=\begin{pmatrix}
2&-i&0&0\\
i&2&0&0\\
0&0&-2&-i\\
0&0&i&-2
\end{pmatrix},\\
\rho_2=\begin{pmatrix}
0&-\sqrt{3}&-1&i\\
-\sqrt{3}&0&i&1\\
-1&-i&0&-\sqrt{3}\\
-i&1&-\sqrt{3}&0
\end{pmatrix},
\end{array}
\end{align}

\begin{align}
\begin{array}{c}
S_{T,1}=\begin{pmatrix}
0&-\sqrt{3}&0&0\\
-\sqrt{3}&0&0&0\\
0&0&0&\sqrt{3}\\
0&0&\sqrt{3}&0
\end{pmatrix},\
S_{T,3}=\begin{pmatrix}
0&0&\sqrt{3}&0\\
0&0&0&\sqrt{3}\\
\sqrt{3}&0&0&0\\
0&\sqrt{3}&0&0
\end{pmatrix},\\
\\
S_{T,2}=\mathrm{diag.}(-\sqrt{3},\sqrt{3},\sqrt{3},-\sqrt{3})
\end{array}
\end{align}

\begin{align}
\begin{array}{c}
S_{O,1}=\begin{pmatrix}
-\sqrt{3}&0&4i&-1\\
0&\sqrt{3}&-1&-4i\\
-4i&-1&-\sqrt{3}&0\\
-1&4i&0&\sqrt{3}
\end{pmatrix},\quad
S_{O,3}=\begin{pmatrix}
2&4i&0&0\\
-4i&2&0&0\\
0&0&-2&4i\\
0&0&-4i&-2
\end{pmatrix},\\
S_{O,2}=\begin{pmatrix}
0&-\sqrt{3}&-1&-4i\\
-\sqrt{3}&0&-4i&1\\
-1&4i&0&-\sqrt{3}\\
4i&1&-\sqrt{3}&0
\end{pmatrix}.
\end{array}
\end{align}
Then we find the algebra is
\begin{align}
[\rho_1,\rho_2]&=2i\rho_3\\
[S_{T,1},S_{T,2}]&=-i\frac{6}{5}\rho_3+i\frac{6}{5}S_{O,3}\\
[S_{O,1},S_{O,2}]&=-12i\rho_3+4iS_{O,3}\\
[S_{T,1},\rho_2]+[\rho_1,S_{T,2}]&=-4iS_{T,3}\\
[S_{O,1},\rho_2]+[\rho_1,S_{O,2}]&=-6iS_{O,3}\\
[S_{O,1},S_{T,2}]+[S_{T,1},S_{O,2}]&=16iS_{T,3},
\end{align}
and their cyclic permutations of $1,2$ and $3$.
Note that the algebra of $\rho_j$ and $S_{O,\,j}$ closes themselves, whereas that of $\rho_j$ and $S_{T,\,j}$ does not.
This is the reason that the 4-caloron and also the 4-monopole with tetrahedral symmetry 
do not exist alone: they only arise in the case with  an extreme limit of the parameter, as we will see shortly.

The bulk Nahm data are given in \cite{HoughtonSutcliffe1996CMP} as
\begin{align}
T_j(s)=x(s)\;\rho_j+y(s)\;S_{T,\,j}+w(s)\;S_{O,\,j},\label{OctaTetra-4 bulk data}
\end{align}
with 
\begin{align}
x(s)&=-\frac{\omega}{10}\left(-4\sqrt{\wp(u)}+\frac{\wp'(u)}{2\wp(u)}\right),\label{4-caloron x}\\ 
y(s)&=-\omega\frac{a}{2\wp(u)},\label{4-caloron y}\\ 
w(s)&=\frac{\omega}{10}\left(\sqrt{\wp(u)}+\frac{\wp'(u)}{2\wp(u)}\right), \label{4-caloron w}
\end{align}
where $u=\omega(s+1)$ and $a$ is a real parameter. 
In this case, the Weierstrass elliptic funciton solves
\begin{align}
\left(\wp'(u)\right)^2=4\wp^3(u)-4\wp(u)+12a^2,
\end{align}
with the half-peiriod
\begin{align}
\omega=\frac{1}{\sqrt{e_1-e_2}}K\left(\sqrt{\frac{e_3-e_2}{e_1-e_2}}\right),\label{TO4 omega}
\end{align}
where
\begin{align}
e_1&=\frac{1}{\sqrt[3]{18}}\;b+\sqrt[3]{\frac{2}{3}}\;\frac{1}{b}\ ,\label{TO4 e1}\\
e_2&=-\frac{1}{\sqrt[3]{18}}e^{-\pi i/3}\;b-\sqrt[3]{\frac{2}{3}}e^{\pi i/3}\;\frac{1}{b}\ ,\label{TO4 e2}\\
e_3&=-\frac{1}{\sqrt[3]{18}}e^{\pi i/3}\;b-\sqrt[3]{\frac{2}{3}}e^{-\pi i/3}\;\frac{1}{b}\  ,\label{TO4 e3}\\
b&=\left(\sqrt{3(243a^4-4)}-27a^2\right)^{1/3},\label{TO4 b}
\end{align}
and $K$ is the complete elliptic integral of the first kind.
The parameter takes values in the range $|a|<3^{-5/4}\sqrt{2}\simeq 0.35819$ due to the reason from the analysis of the half-period $\omega$ \cite{HoughtonSutcliffe1996CMP}.

Note that when the parameter $a=0$, the bulk data turns out to be that of the octahedral 4-monopole
given in \cite{HMM1995}, which is also utilized in \cite{Ward2004} as the bulk data of the octahedral 4-caloron.
Although  the apparent form of the bulk Nahm data with $a=0$ is different from that of \cite{Ward2004,HMM1995}, 
 these are equivalent  by means of the addition theorem of the Weierstrass elliptic function,
see Appendix.

\begin{figure}[htbp]
\begin{center}
\begin{minipage}{0.45\hsize}
\begin{center}
(a) $a=0.3$

 \includegraphics[width=45mm]{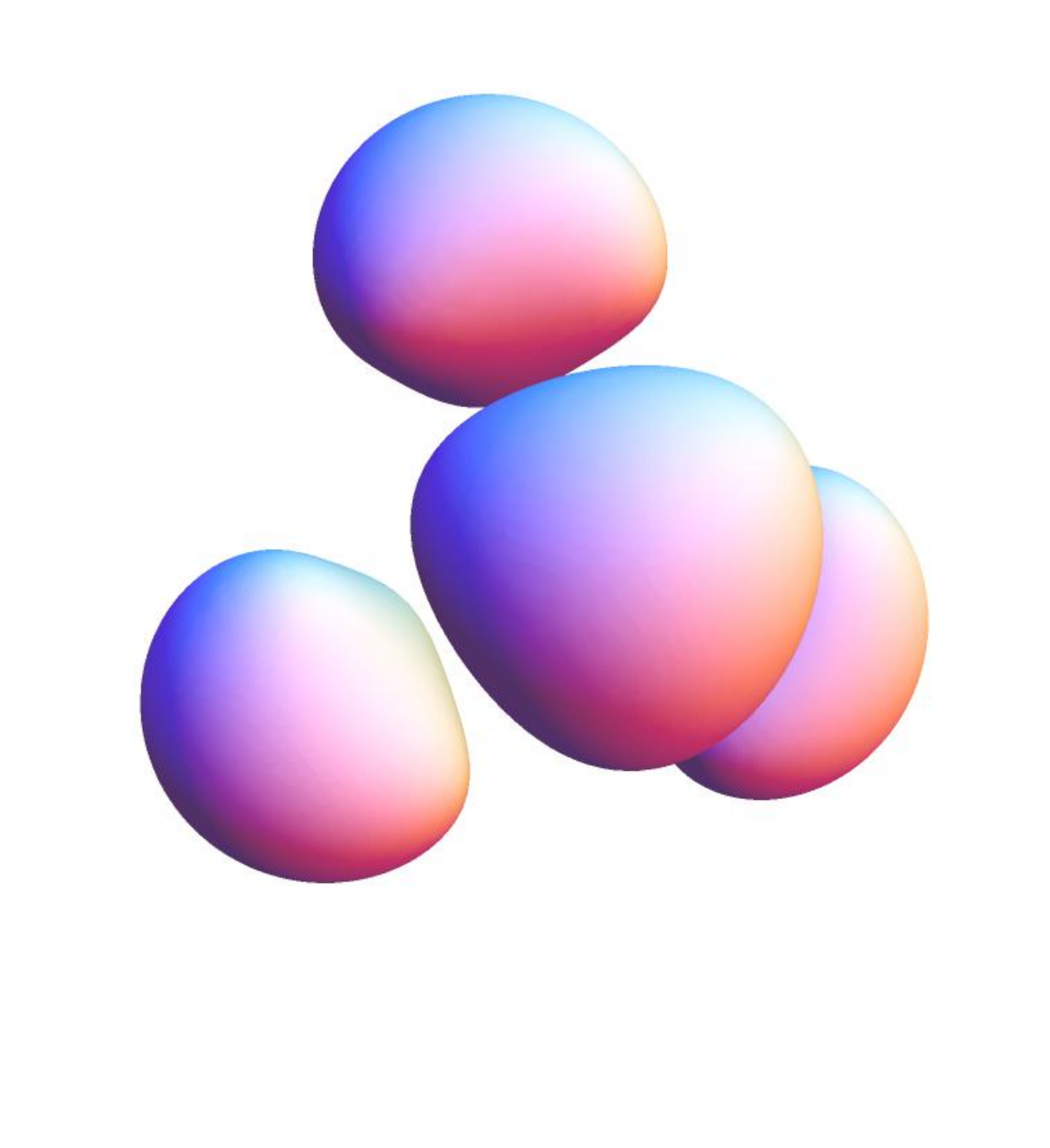}
\end{center}
\end{minipage}
\begin{minipage}{0.45\hsize}
\begin{center}
(b) $a=0.15$

\includegraphics[width=45mm]{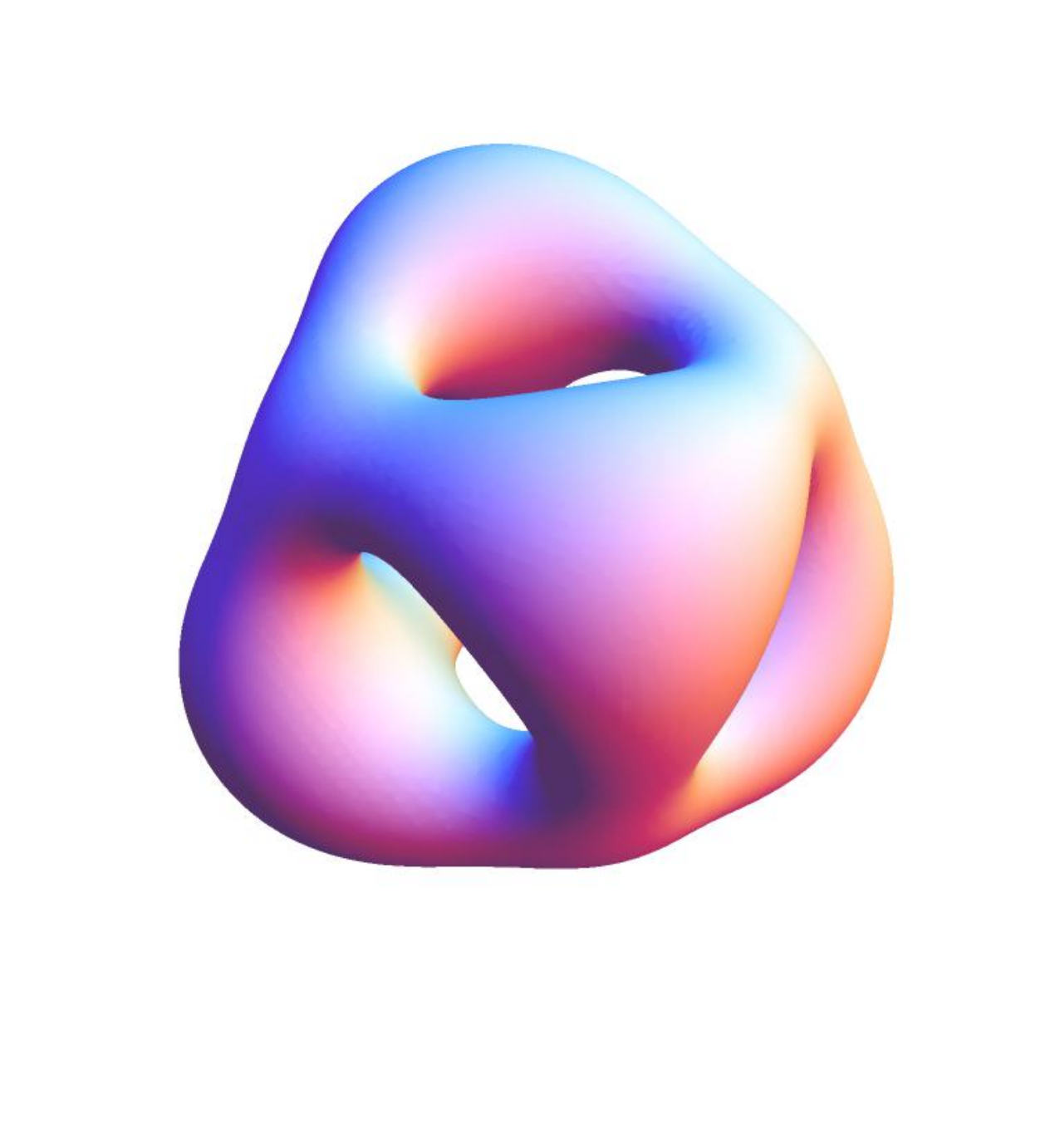}
\end{center}
\end{minipage}

\begin{center}
\begin{minipage}{0.45\hsize}
\begin{center}
(c) $a=0$

\includegraphics[width=45mm]{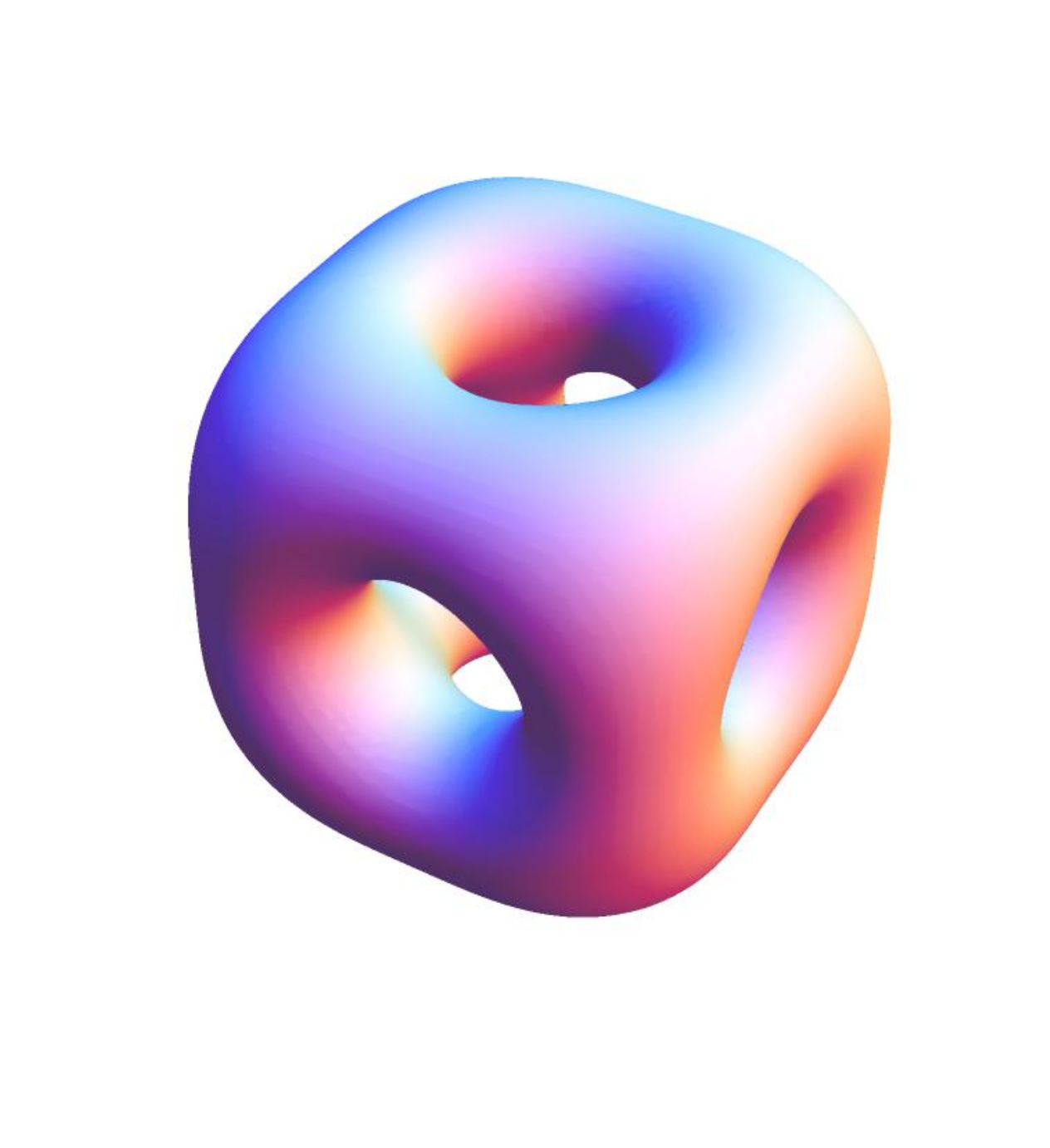}
\end{center}
\end{minipage}
\end{center}

\begin{minipage}{0.45\hsize}
\begin{center}
(d) $a=-0.15$

\includegraphics[width=45mm]{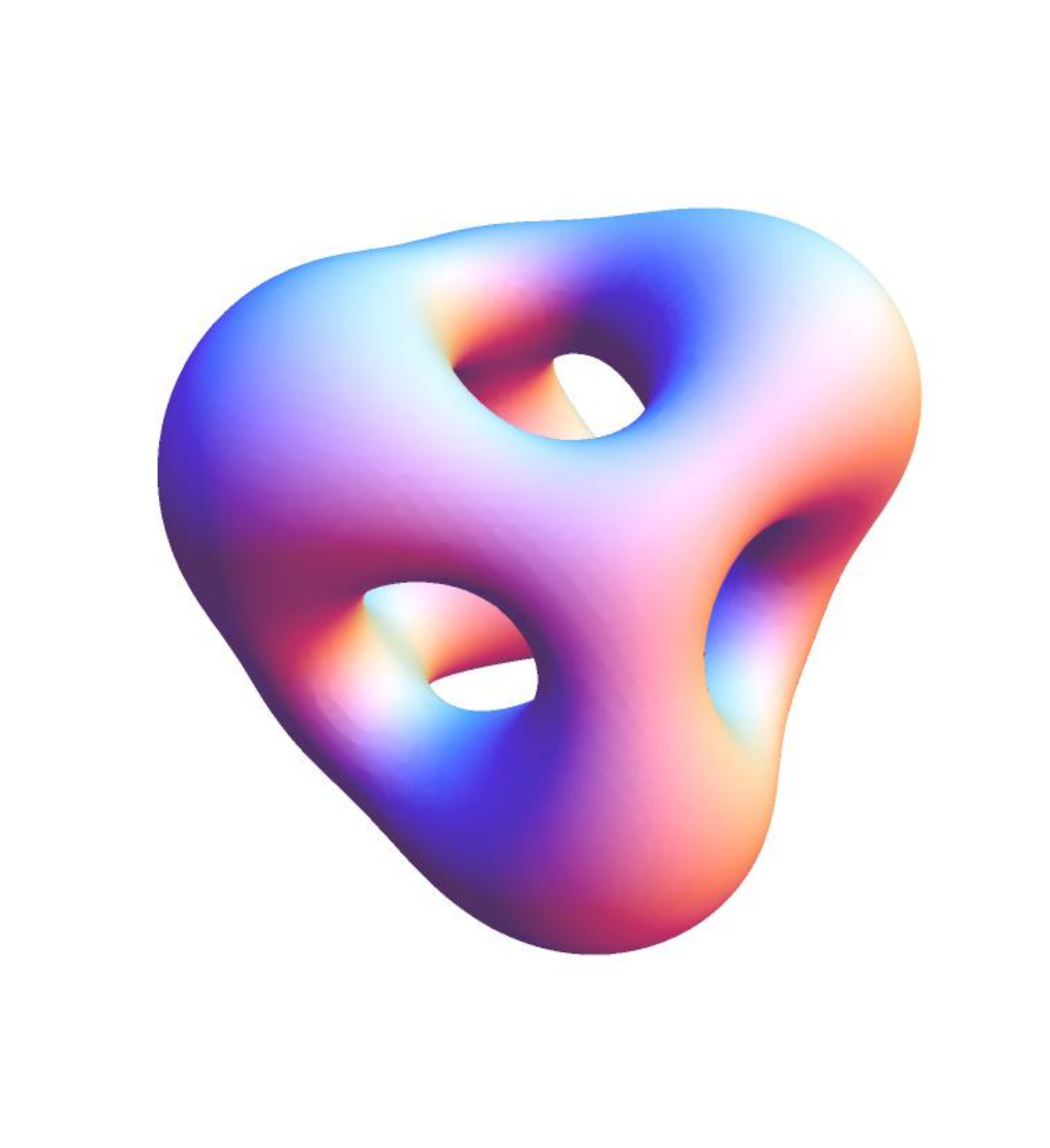}
\end{center}
\end{minipage}
\begin{minipage}{0.45\hsize}
\begin{center}
(e) $a=-0.3$

\includegraphics[width=45mm]{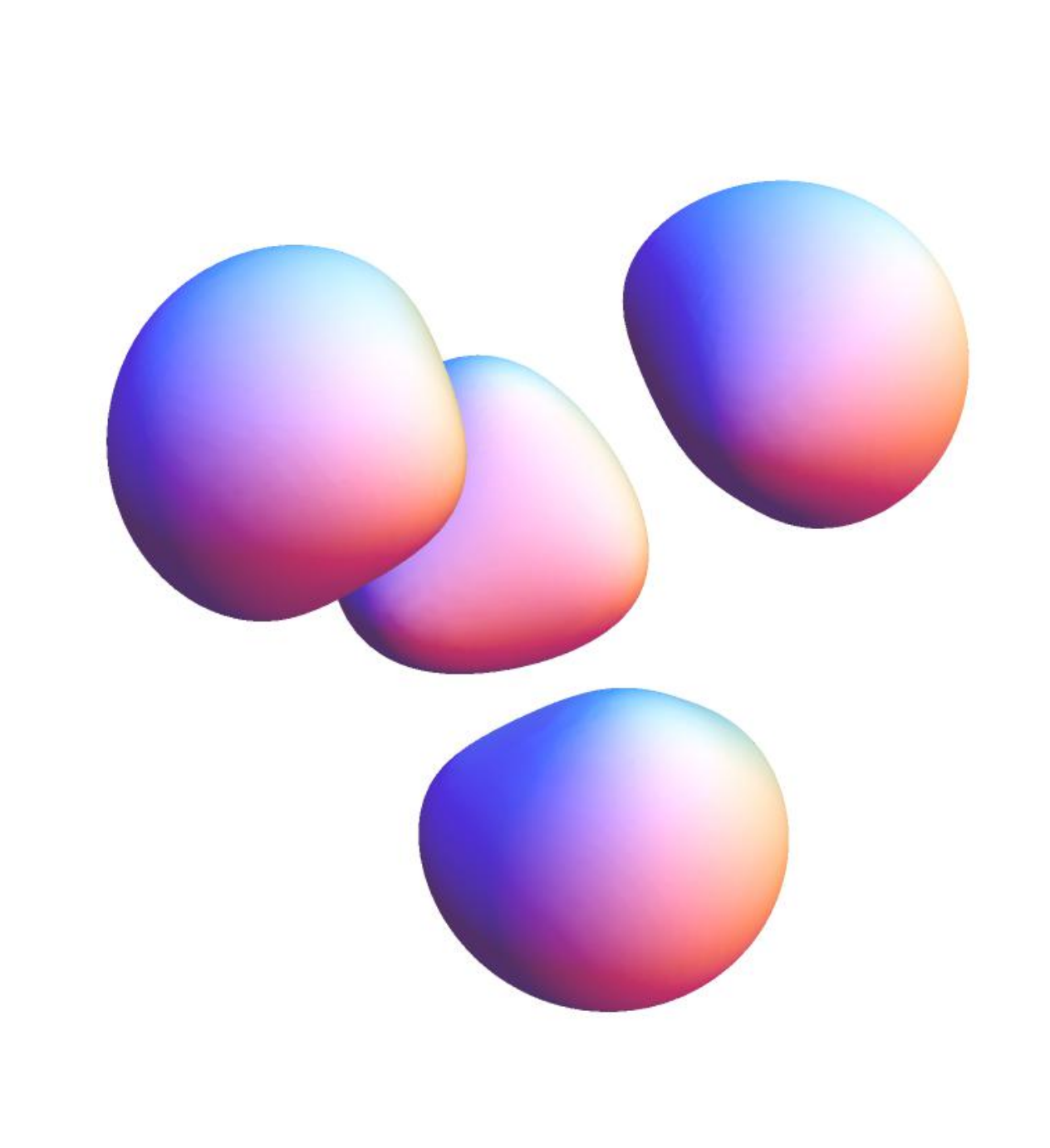}
\end{center}
\end{minipage}
\caption{Equi-action density surfaces plot for the 4-caloron with one parameter in $\mathbb{R}^3$ at $x^0=0$, with several parameter values. 
All of the action density values are $0.7$.}\label{TO4 caloron}
\end{center}
\end{figure}

The solution to the boundary Nahm equation (\ref{Boundary Nahm}) is
\begin{align}
W=\lambda\left(\mathbf{1}_2,-i\sigma_3,-i\sigma_1,i\sigma_2\right)
\end{align}
where $\lambda=\sqrt{-2x(\mu_0/2)+8w(\mu_0/2)}$.
Having obtained the bulk and boundary Nahm data, we perform the numerical Nahm transform to observe
the action density in the configuration space at several  values of $a$.
In Figure \ref{TO4 caloron},  we find the ``4-caloron scattering" configurations as the parameter varies, where $\mu_0=1.0$ is applied.

\begin{figure}[htbp]
\begin{center}
\begin{minipage}{0.45\hsize}
\begin{center}
(a) $a=0.35$

 \includegraphics[width=45mm]{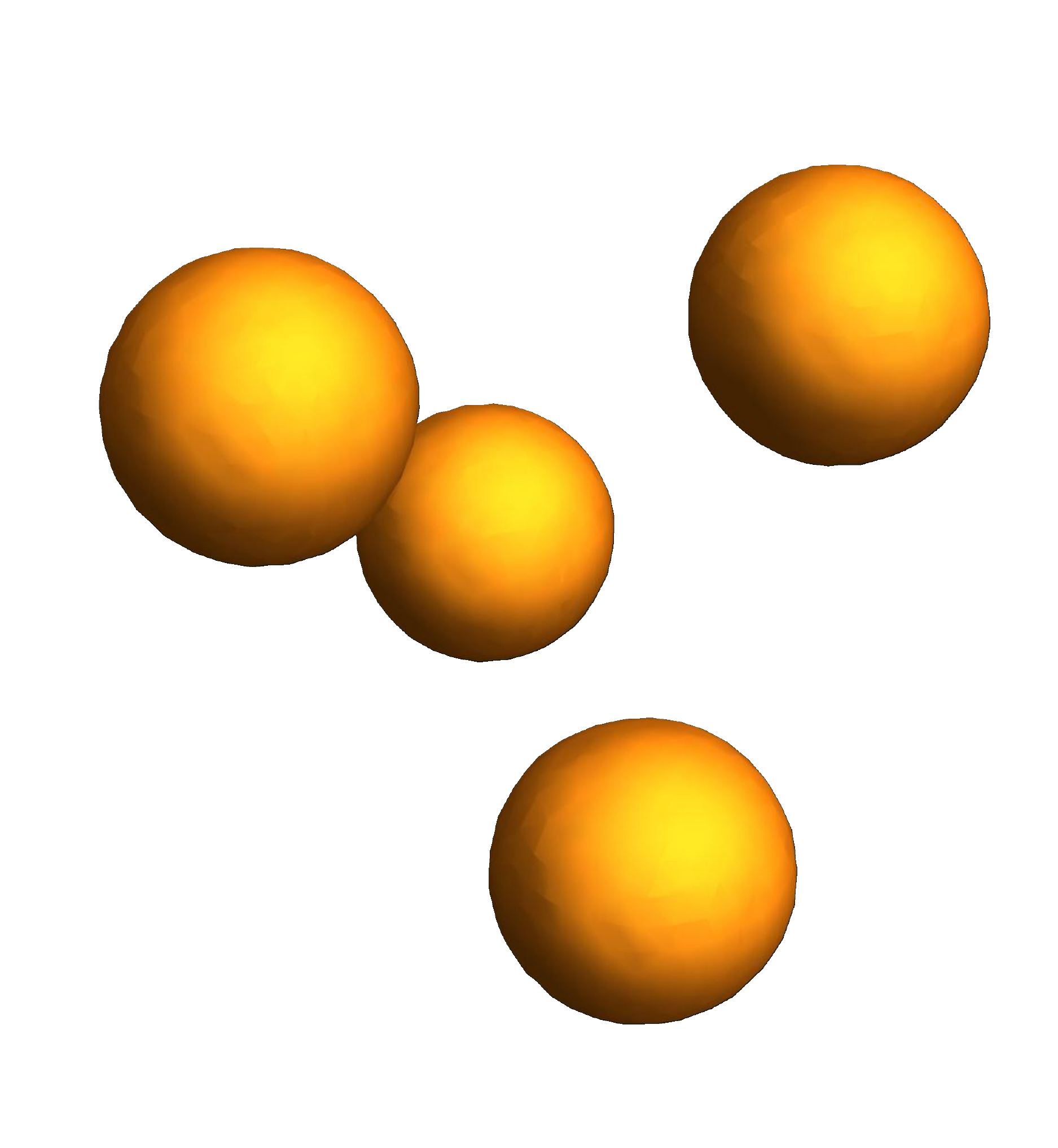}
\end{center}
\end{minipage}
\begin{minipage}{0.45\hsize}
\begin{center}
(b) $a=0.15$

\includegraphics[width=45mm]{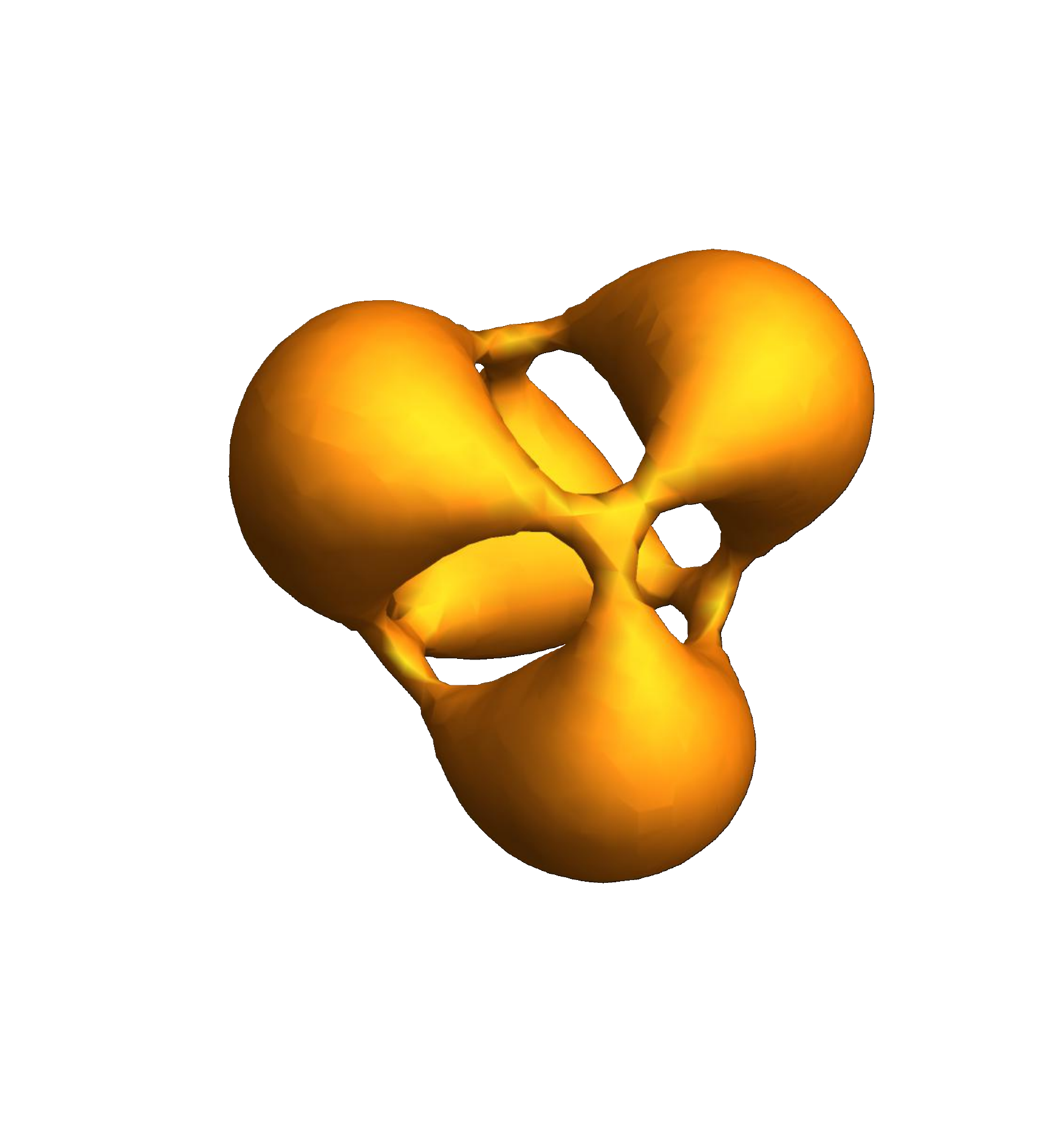}
\end{center}
\end{minipage}

\begin{center}
\begin{minipage}{0.45\hsize}
\begin{center}
(c) $a=0$

\includegraphics[width=45mm]{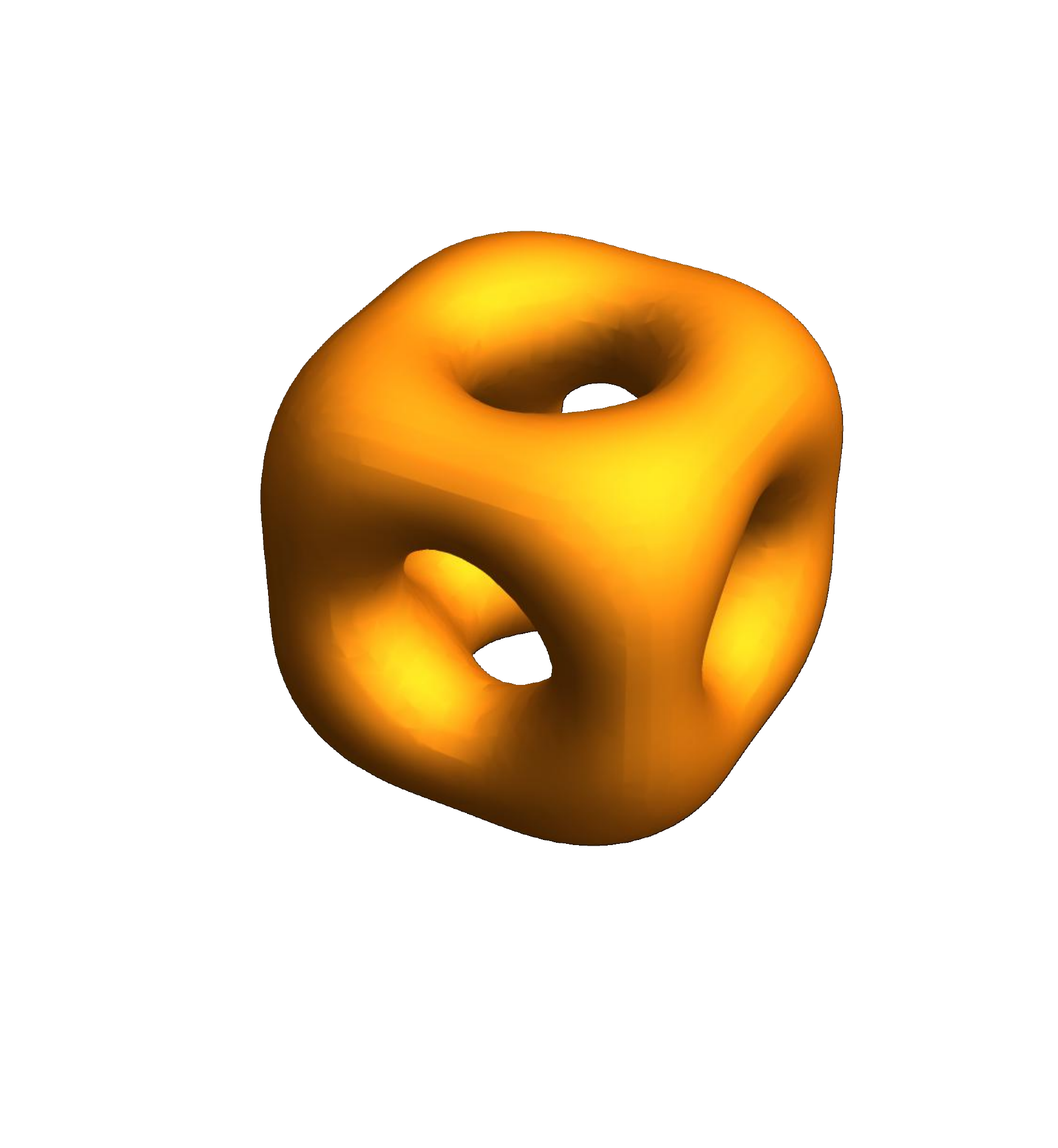}
\end{center}
\end{minipage}
\end{center}

\begin{minipage}{0.45\hsize}
\begin{center}
(d) $a=-0.15$

\includegraphics[width=45mm]{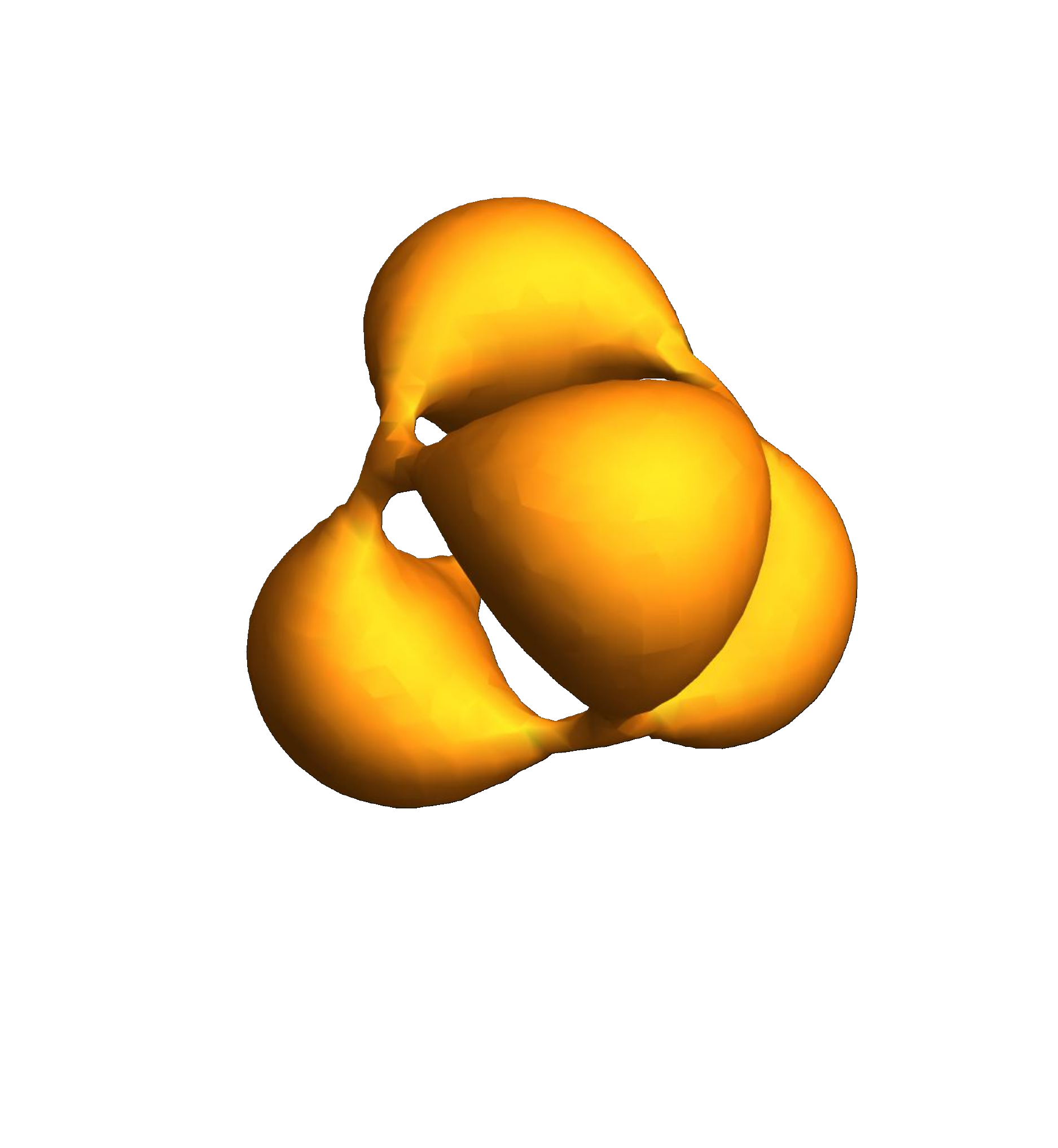}
\end{center}
\end{minipage}
\begin{minipage}{0.45\hsize}
\begin{center}
(e) $a=-0.35$

\includegraphics[width=45mm]{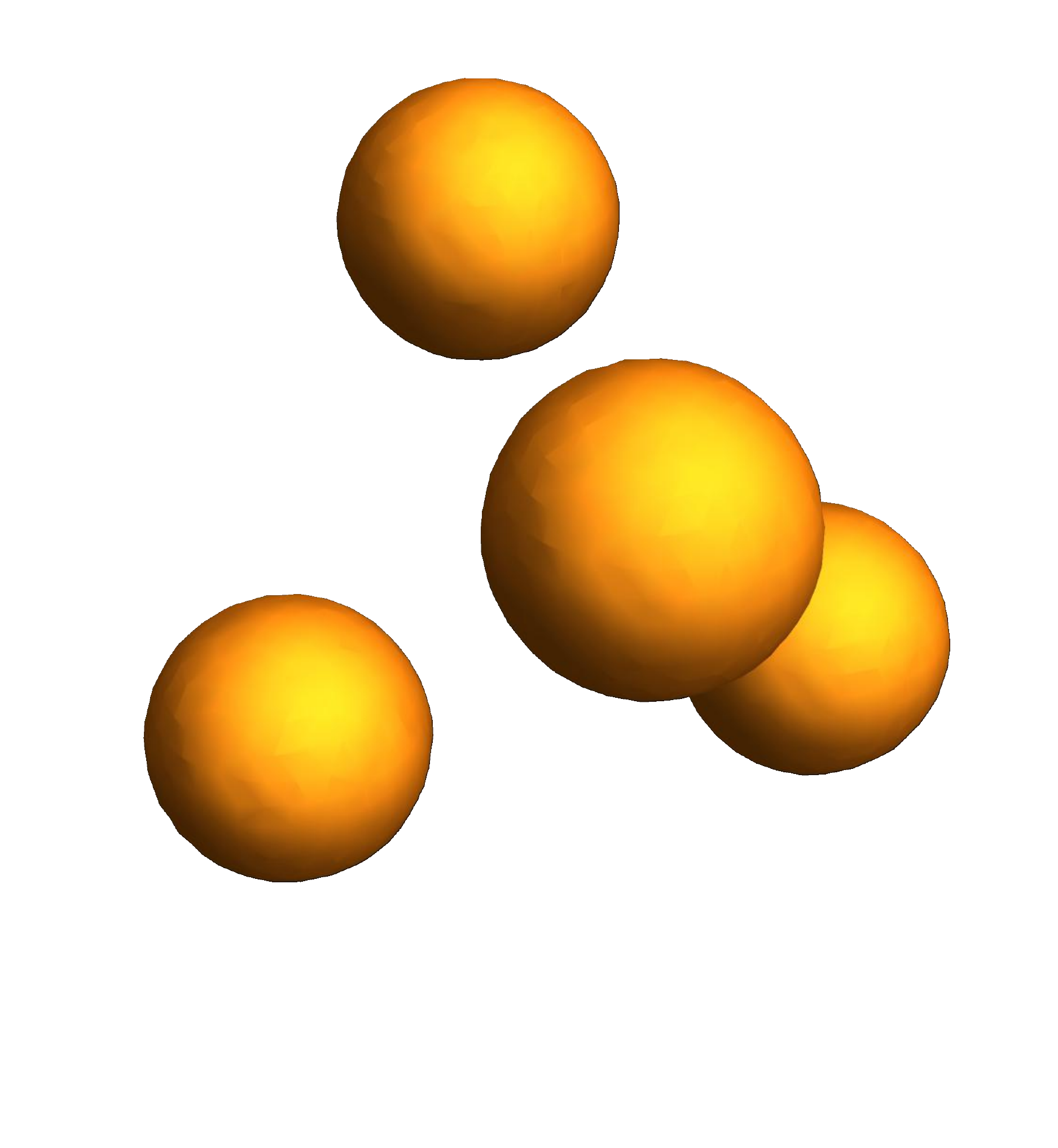}
\end{center}
\end{minipage}
\caption{Equi-action density surfaces plot for the 4-instanton limit of the 4-caloron with one parameter in $\mathbb{R}^3$ at $x^0=0$, with several parameter values. 
All of the values $s_{\mathrm{inst}}=2.8$.
The relative scale and the orientation with respect to Figure \ref{TO4 caloron} are not significant.}\label{TO4 instanton plot}
\end{center}
\end{figure}

Finally, we consider the large period limit of the 4-caloron with parameter $a$.
The ordinary procedure leads to the following ADHM data from the Nahm data obtained above,
\begin{align}
\Delta=\omega
\begin{pmatrix}
\tilde{W}\\
 B_1+B_2
\end{pmatrix}
+\begin{pmatrix}
0&0&0&0\\
x&0&0&0\\
0&x&0&0\\
0&0&x&0\\
0&0&0&x
\end{pmatrix}
\end{align}
where
\begin{align}
B_1&=\kappa_1
\begin{pmatrix}
2i\sigma_3-\sqrt{3}i\sigma_1 &-\sqrt{3}i\sigma_2 & -i\sigma_2& -i\sigma_1\\
-\sqrt{3}i\sigma_2 &\sqrt{3}i\sigma_1+2i\sigma_3&-i\sigma_1&i\sigma_2\\
-i\sigma_2 &-i\sigma_1&  -\sqrt{3}i\sigma_1-2i\sigma_3 & -\sqrt{3}i\sigma_2\\
-i\sigma_1 & i\sigma_2 & -\sqrt{3}i\sigma_2  &\sqrt{3}i\sigma_1-2i\sigma_3\\
\end{pmatrix},\\
B_2&=\kappa_2\begin{pmatrix}
i\sigma_2 &i\sigma_1 & -i\sigma_3 &0\\
i\sigma_1 &-i\sigma_2 &0& -i\sigma_3 \\
-i\sigma_3&0 &-i\sigma_2 & -i\sigma_1 \\
0&-i\sigma_3 &-i\sigma_1 & i\sigma_2 
\end{pmatrix},
\end{align}
and
\begin{align}
\tilde{W}=\kappa_3\left(\mathbf{1}_2,-i\sigma_3,-i\sigma_1,i\sigma_2\right).
\end{align}
Here,  $\kappa_1=\sqrt{e_1}/2,\; \kappa_2=\sqrt{3}a/2e_1$ 
and $ \kappa_3=\sqrt{(3e_1^2-1)/2e_1}$.
The equi-action density surface plots in $\mathbb{R}^3$ are similar to that of the caloron in analysis, as shown in Figure \ref{TO4 instanton plot}.

We therefore find that the 4-monopole interpolating between tetrahedral and cubic shape can smoothly be cooled down to instanton through the corresponding 4-caloron.

\newpage

\subsection{Non-Existence of 6-caloron with Icosahedral Symmetry}
In this subsection, we consider the 6-caloron with icosahedral symmetry.
It is shown in \cite{HMM1995} that the Nahm data of 6-momopole with similar symmetry does not exist,
due to the fact that there is no solution to enjoy the boundary condition for the monopoles.
Namely, the residue at the boundaries does not define an irreducible representation of $su(2)$.
In contrast, it is not necessary the monopole boundary conditions for the caloron Nahm data: we need the boundary 
conditions (\ref{Boundary Nahm}) instead.
Hence there remains possibility that  the Nahm data of 6-caloron with icosahedral symmetry exists.
We show, however, the non-existence of such Nahm data in the following, as well as the monopoles.
 
For the construction of the bulk Nahm data of 6-caloron with icosahedral symmetry, we apply the 
following representation of $so(3)$, $\rho_j$, and the icosahedral invariant vector $S_j$ \cite{HMM1995},
\begin{align}
\rho_1&=
\begin{pmatrix}
0&\sqrt{5}&0&0&0&0\\
\sqrt{5}&0&2\sqrt{2}&0&0&0\\
0&2\sqrt{2}&0&3&0&0\\
0&0&3&0&2\sqrt{2}&0\\
0&0&0&2\sqrt{2}&0&\sqrt{5}\\
0&0&0&0&\sqrt{5}&0
\end{pmatrix},\\
\rho_2&=i
\begin{pmatrix}
0&-\sqrt{5}&0&0&0&0\\
\sqrt{5}&0&-2\sqrt{2}&0&0&0\\
0&2\sqrt{2}&0&-3&0&0\\
0&0&3&0&-2\sqrt{2}&0\\
0&0&0&2\sqrt{2}&0&-\sqrt{5}\\
0&0&0&0&\sqrt{5}&0
\end{pmatrix},\\
\rho_3&=\mathrm{diag.}\;(5,3,1,-1,-3,-5)
\end{align}
and
\begin{align}
S_1&=
\begin{pmatrix}
0&\sqrt{5}&0&0&-7\sqrt{5}&0\\
\sqrt{5}&0&-5\sqrt{2}&0&0&7\sqrt{5}\\
0&-5\sqrt{2}&0&10&0&0\\
0&0&10&0&-5\sqrt{2}&0\\
-7\sqrt{5}&0&0&-5\sqrt{2}&0&\sqrt{5}\\
0&7\sqrt{5}&0&0&\sqrt{5}&0
\end{pmatrix},\\
S_2&=i
\begin{pmatrix}
0&-\sqrt{5}&0&0&-7\sqrt{5}&0\\
\sqrt{5}&0&5\sqrt{2}&0&0&7\sqrt{5}\\
0&-5\sqrt{2}&0&-10&0&0\\
0&0&10&0&5\sqrt{2}&0\\
7\sqrt{5}&0&0&-5\sqrt{2}&0&-\sqrt{5}\\
0&-7\sqrt{5}&0&0&\sqrt{5}&0
\end{pmatrix},\\
S_3&=\begin{pmatrix}
-2&0&0&0&0&-14\\
0&10&0&0&0&0\\
0&0&-20&0&0&0\\
0&0&0&20&0&0\\
0&0&0&0&-10&0\\
-14&0&0&0&0&2
\end{pmatrix}.
\end{align}
Then we obtain the algebra of these representations,
\begin{align}
[\rho_1,\rho_2]&=2i\rho_3\\
[S_1,S_2]&=-100i\rho_3-10iS_{3}\\
[S_1,\rho_2]+[\rho_1,S_2]&=-10iS_3,
\end{align}
and cyclic permutations of $1,2$ and $3$.
As in the previous cases, the bulk Nahm data is defined as
\begin{align}
T_j(s)=x(s)\rho_j+y(s)S_j,
\end{align}
and the solution is 
\begin{align}
x(s)&=-\frac{5\kappa}{6}\left[\frac{\wp'(u)}{\wp(u)}+\frac{3\wp(u)^2}{\wp'(u)}\right],\label{6-caloron x}\\
y(s)&=-\frac{\kappa}{\wp(u)\wp'(u)},\label{6-caloron y}
\end{align}
where $u=(35\kappa s/3)+u_0$, and the Weierstrass function solves
\begin{align}
\left(\wp'(u)\right)^2=4\wp^3(u)-4.
\end{align}
Here the constant $\kappa$ and $u_0$ will be chosen in accordance with the periodicity of the bulk data.
We find from (\ref{6-caloron x}) and (\ref{6-caloron y}) that the bulk data have poles at the half-period of the Weierstrass function $u=\omega \ (\mathrm{mod}\; 2\omega)$ in addition to $u=0 \ (\mathrm{mod}\; 2\omega)$.
Hence we  should take the defining region of the bulk data as $u\in(0,\omega)$, unlike the other cases considered previously. 
For example, if we take the periodicity on the real axis of $u$, that is, $u|_{s=-1}=0$ and $u|_{s=1}=\omega$,
where 
\begin{align}
\omega=\frac{1}{4\sqrt{3\pi}}\;\Gamma\left(1/3\right)\Gamma\left(1/6\right)
\end{align}
is the half period along the real axis,
then we find $\kappa=3\omega/70$ and $u_0=\omega/2$.
Thus, the bulk data in this case turns out to be
\begin{align}
x(s)&=\frac{\omega}{28}\left[\frac{\wp'(u)}{\wp(u)}+\frac{3\wp(u)^2}{\wp'(u)}\right],\label{6-x bulk data}\\
y(s)&=\frac{3\omega}{70 \wp(u)\wp'(u)}\label{6-y bulk data},
\end{align}
with $u=\omega(s+1)/2$.
Note that we have other choices of $u$ in the complex $u-$plane, for example, $u=\tilde{\omega}(s+1)/3$,
where 
\begin{align}
\tilde\omega=\frac{i}{4\sqrt{\pi}}\;\Gamma\left(1/3\right)\Gamma\left(1/6\right)
\end{align}
 is another half-period of the elliptic function.
In the following consideration, however, 
 we apply only the first case, because the analysis is similar in any case.

\begin{figure}[h]
\begin{center}
\includegraphics[width=70mm]{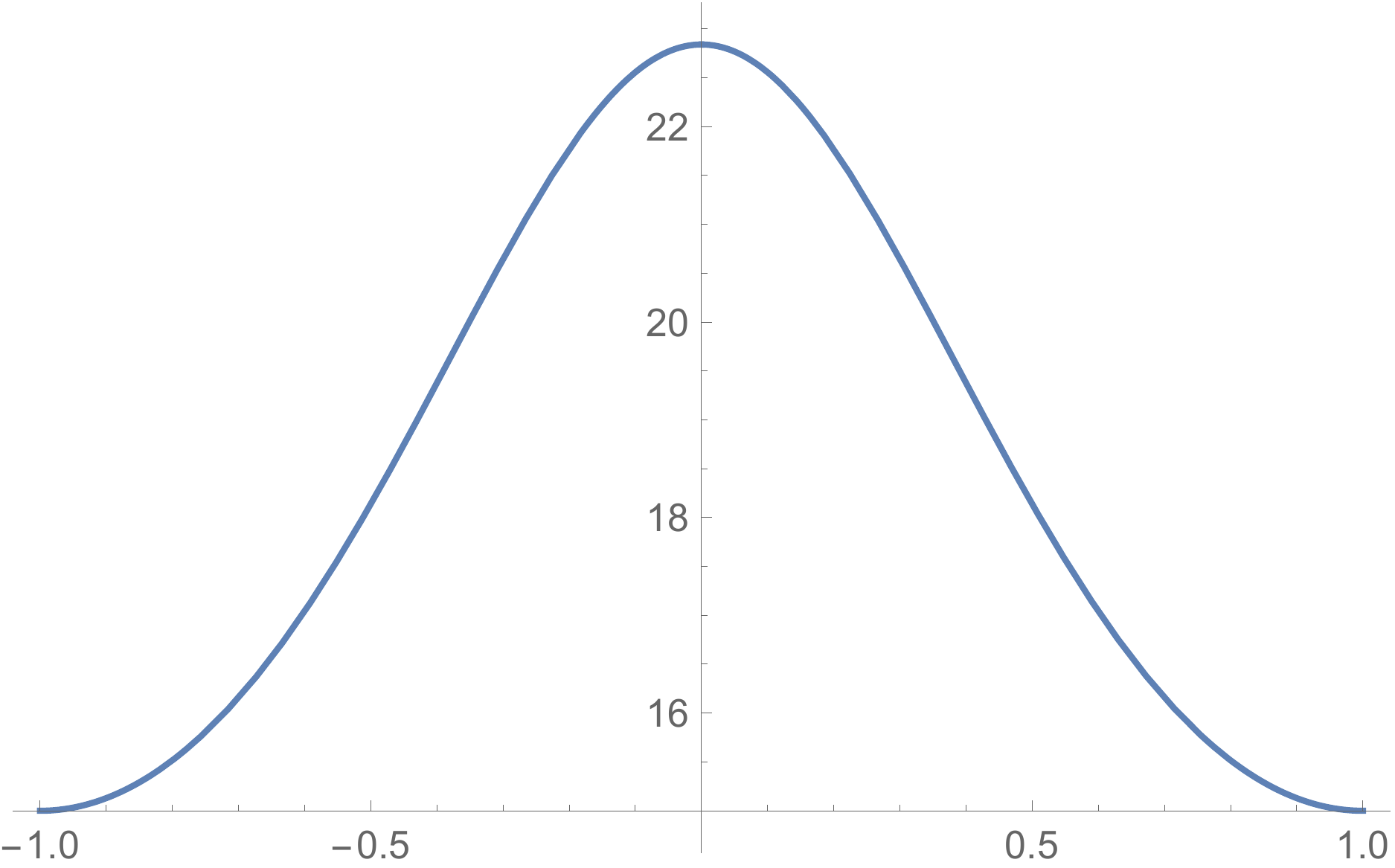}
\includegraphics[width=70mm]{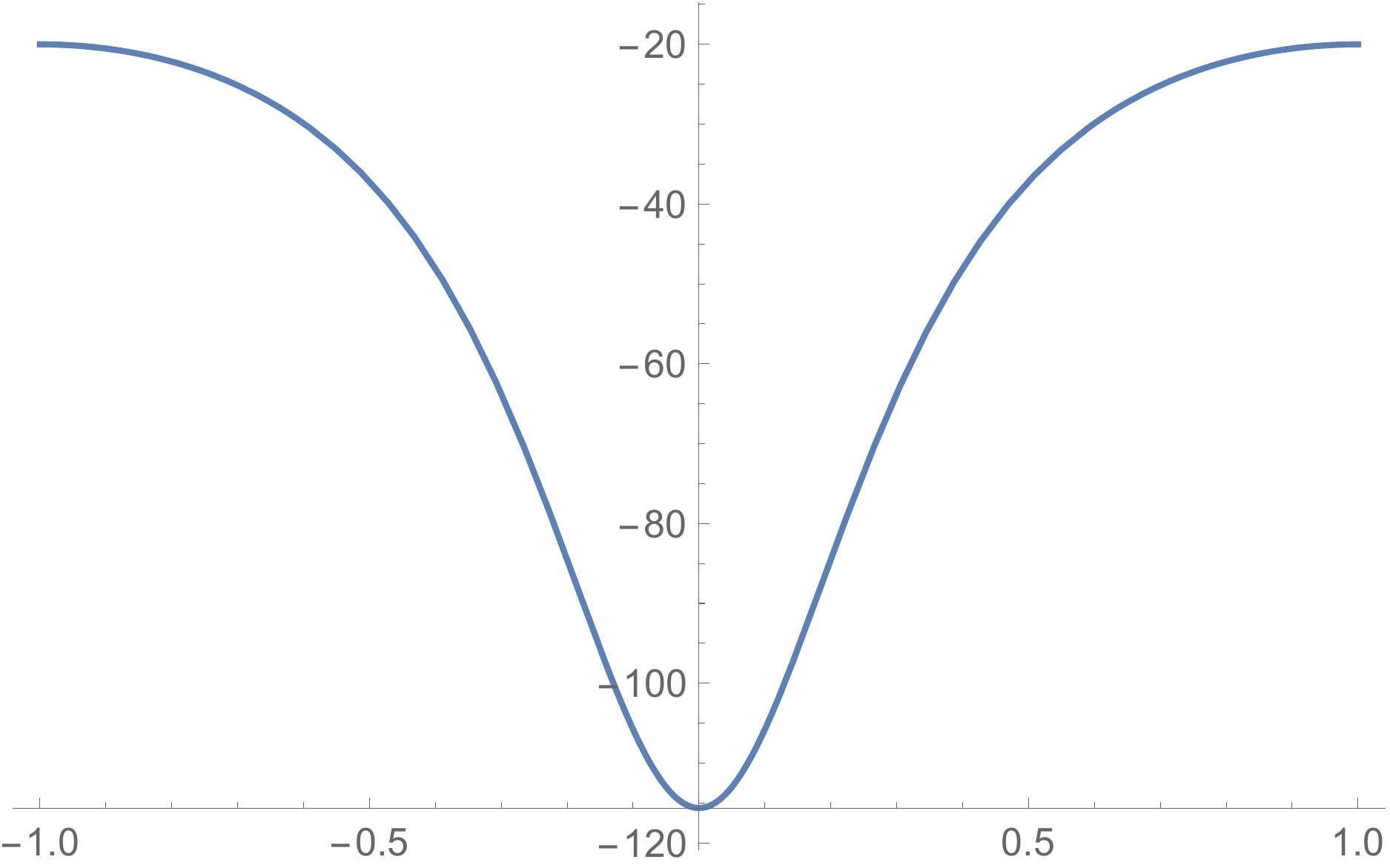}
\caption{Plots of  $\left\{x(s)-x(-s)\right\}/\left\{y(s)-y(-s)\right\}$ (left), 
and $\left\{x(s)+x(-s)\right\}/\left\{y(s)+y(-s)\right\}$ (right) from the bulk Nahm data  (\ref{6-x bulk data}) and (\ref{6-y bulk data}).
It is obvious that they are not constant functions.}\label{non-existence of 6-caloron}
\end{center}
\end{figure}

We now investigate whether it is possible to verify the reality conditions $T_j(-s)={}^tT_j(s)$, 
in addition to the hermiticity $T_j^\dag(s)=T_j(s)$ from the bulk data obtained above.
In order to fulfill these conditions, it is necessary to find a reality basis of the matrices $\rho_j$ and $S_j$ through a unitary transformation.
As in the previous cases, the functions $x(s)$ and $y(s)$ are real valued, thus,
after the unitary transformation, the diagonal components, and the symmetric part of the off-diagonal components 
of $\rho_j$ and $S_j$ have to be real; on the other hand, the skew-symmetric part of them have to be pure imaginary.
Hence we observe that the components of $T_j(s)$, which are proportional to the linear combination of $x(s)$ and $y(s)$, have to be real and even function with respect to $s$ for the diagonal and the symmetric part, and pure imaginary and odd for the skew-symmetric part, respectively.
We therefore observe that there have to be, at least, constants such that the linear combination of $x(s)$ and 
$y(s)$ becomes even and odd functions.
Namely,   we need to confirm the existence of constants $\alpha$ and $\beta$ with 
$x(s)+\alpha y(s)=x(-s)+\alpha y(-s)$ and $x(s)+\beta y(s)=-x(-s)-\beta y(-s)$.
These are necessary conditions for the bulk Nahm data, which are satisfied by all of the other caloron Nahm data considered in this section, so far.
However, we find from Figure \ref{non-existence of 6-caloron} that
there are no such constants from the bulk data (\ref{6-x bulk data}) and (\ref{6-y bulk data}).

As mentioned, there are other choices of the periodicity in the complex $u$-plane for the functions (\ref{6-x bulk data}) and (\ref{6-y bulk data}) as the bulk Nahm data.
However, the situation is similar for any choice of the period with the investigation given above.
We therefore conclude that there is no bulk Nahm data for the 6-caloron with icosahedral symmetry.

As mentioned,  we have been able to confirm the existence of such
 constants for the cases considered in the previous subsections, so that the unitary transformation into the reality base can be found.
In the case of the 4-caloron with one-parameter, there have been three functions $x(s), y(s)$ and $w(s)$ in 
the bulk Nahm data.
However, the function $y(s)$ is already even itself, thus the situation is to construct an even and odd functions from the linear combination of $x(s)$ and $w(s)$, which is similar to the other cases.

\section{Summary and Outlook}

In this paper, we have considered the  higher charge calorons with particular spatial symmetries of polyhedral shapes.
Those are, the 5-caloron with octahedral symmetry, the 7-caloron with icosahedral symmetry, and the 
4-caloron interconnecting octahedral and tetrahedtal symmetries, respectively, together with their large period limits.
Having obtained the analytic Nahm data for each object,
we have also performed the numerical Nahm transform for visualizing of their structures in configuration space \textit{\`a la} \cite{MNST}.
In addition, we show the non-existence of the  6-caloron with icosahedral symmetry from the investigation of the Nahm data.

For the future work, as mentioned in Introduction,  
discovering the more general cases of the calorons with and without specific symmetry will be helpful to understand the 
whole structure of Yang-Mills related solitons.
For example, it would be interesting to construct the symmetric calorons with instanton charge beyond 7,  with moduli parameters, \etc.
Particular interest is the calorons with non-trivial holonomy.
The calorons considered in this paper have been restricted to trivial holonomy cases, in other words, exactly periodic ones.
The calorons with non-trivial holonomy were firstly discovered by Kraan-van Baal \cite{KvB} and Lee-Lu \cite{LL} (KvBLL). 
The KvBLL caloron has a composite structure with two constituent ``monopoles", and is axially symmetric in $\mathbb{R}^3$.
The massless limit of one of the monopole 
reduces to the 1-caloron of trivial holonomy, \ie Harrington-Shepard 1-caloron \cite{HS}.
The class of calorons with non-trivial holonomy are crucial in the analysis for the confinement/deconfinement phase transition of
quarks in QCD \cite{GPY,ShifmanUnsal,Diakonov,DGPS,PoppitzSchaferUnsal,DiaGro,GroSli,Sli,DiakonovPetrov}.
The higher charge generalization for the non-trivial holonomy calorons is quite restricted except for the cases of instanton charge two \cite{BNvB, Harland, Kato2018,NakamulaSakaguchi}.
This situation is reflecting the fact that the introduction of non-trivial holonomy in the Nahm data is
extremely complicated problem, \ie it is hard to construct the corresponding Nahm data.
Even if the bulk Nahm data is known, it is quite non-trivial work to find the boundary Nahm data in general.  
In contrast, as we have shown in this paper,  the monopole Nahm data can be generalized straightforwardly to
the caloron Nahm data with trivial holonomy. 
It is not clear how to incorporate a non-trivial holonomy parameter with a given monopole Nahm
data, as far as the knowledge of the present authors.
Thus, the discovery of the symmetric calorons with non-trivial holonomy accompanied by
monopole charge greater than two is particularly interesting direction of the study of the BPS solitons.

As mentioned earlier, calorons have been used to approximate the gauged Skyrmions \cite{Cork2018} with the Atiyah-Manton-Sutcliffe procedure.
The results suggest there will be a deep connection between the BPS solutions to the Yang-Mills theories and the solutions to the Skyrme-like models.
In particular, quite interesting is the relationship between the symmetric BPS monopoles and the Skyrmions through the rational map constructions \cite{HoughtonMantonSutcliffe1998}, which are also utilized in many cases on the construction of non-BPS solitons, for example, Hopfion solutions to the extended Skyrme models, \eg \cite{GillardSutcliffe2010,Foster2012,BattyeHaberichter2013,SamoilenkaShnir2018}.

Although the whole understanding of the Yang-Mills related solitons is still in progress, the construction of concrete solutions will give a propelling force to insight into the comprehension of the subject.



%

\section*{Appendix}
\renewcommand{\theequation}{A.\arabic{equation}}
In this Appendix, we show the equivalence of the bulk Nahm data of the 4-caloron with octahedral symmetry used in  \cite{Ward2004}, originally 
constructed  in  \cite{HMM1995} for monopole data,  and the bulk data used in section \ref{One parameter family}, which is given primarily in \cite{HoughtonSutcliffe1996CMP}, at the special parameter value $a=0$.

As we have seen in Section \ref{One parameter family}, the bulk Nahm data of the 4-caloron with 
octahedral symmetry is given from (\ref{OctaTetra-4 bulk data}) at $a=0$.
Namely, we find the functions 
(\ref{4-caloron x}), (\ref{4-caloron y}), and (\ref{4-caloron w}) at $a=0$ are reduced to
\begin{align}
x(s)&=-\frac{\omega}{10}\left(-4\sqrt{\wp(u)}+\frac{\wp'(u)}{2\wp(u)}\right),\label{Octa4 x}\\ 
y(s)&=0,\\ 
w(s)&=\frac{\omega}{10}\left(\sqrt{\wp(u)}+\frac{\wp'(u)}{2\wp(u)}\right), \label{Octa4 w}
\end{align}
where the Weierstrass function enjoys $\wp'(u)^2=4\wp(u)^3-4\wp(u)$ with $u=\omega(s+1)$.
We notice that the prime denotes the derivative with respect to the arguments throughout this paper.
The half-period of the Weierstrass function is deduced from (\ref{TO4 omega}--\ref{TO4 b}) at $a=0$.
In this case, we find $e_1=1, e_2=-1$ and $e_3=0$ and consequently 
\begin{align}
\omega=\frac{1}{\sqrt{2\pi}}K\left(1/\sqrt{2}\right)=\frac{1}{4\sqrt{2\pi}}\Gamma\left(1/4\right)^2.\label{Octa4 half-period}
\end{align}
Thus, we find the bulk data concerned is
\begin{align}
T_j(s)=x(s)\;\rho_j+w(s)\;S_{O,\,j},\label{Octa4 bulk data}
\end{align}
with (\ref{Octa4 x}) and (\ref{Octa4 w}).

The other form of bulk Nahm data  has been given by \cite{Ward2004}, originally appeared as 
 the  4-monopole data with octahedral symmety \cite{HMM1995}, \ie
\begin{align}
T^{\mathrm{HMM}}_j(s)=x_{\mathrm{HMM}}(\tilde{u})\;\rho^{\mathrm{HMM}}_j+w_{\mathrm{HMM}}(\tilde{u})\;S^{\mathrm{HMM}}_{O,\,j},\label{Octa4 HMM data}
\end{align}
with 
\begin{align}
x_{\mathrm{HMM}}(\tilde{u})&=-\frac{\tilde{\omega}}{10}\frac{5\wp^2(\tilde{u})-3}{\wp'(\tilde{u})},\label{Octa4 x_HMM}\\ 
w_{\mathrm{HMM}}(\tilde{u})&=\frac{\tilde{\omega}}{10}\frac{1}{\wp'(\tilde{u})}, \label{Octa4 w_HMM}
\end{align}
where $\tilde{u}=\tilde{\omega}(s+1)/2$, and the Weierstrass function enjoys the same differential equation as in the former case, however, 
with the half-period $\tilde{\omega}=(1+i)\omega$ being applied.
Note that the functions $x_{\mathrm{HMM}}$ and $w_{\mathrm{HMM}}$ are originally denoted as $x$ and $y$, respectively, in \cite{HMM1995}.
In  (\ref{Octa4 HMM data}), the generators of $so(3)$ are defined as in (\ref{Octa4 bulk data}), 
$\rho^{\mathrm{HMM}}_j=\rho_j$, whereas the octahedrally invariant vectors differ by factor 2 such as
$S^{\mathrm{HMM}}_{O,\,j}=2S_{O,\,j}$.
Thus, we will show the equivalence $x(u)=x_{\mathrm{HMM}}(\tilde{u})$ and $w(u)=2w_{\mathrm{HMM}}(\tilde{u})$.

First of all, we observe the fact that $\displaystyle 5\wp^2-3=\left(5\wp'^2/4\wp\right)+2$ from the differential equation for $\wp$, then we find
\begin{align}
x_{\mathrm{HMM}}(\tilde{u})=\frac{\tilde{\omega}}{10}\left(\frac{5\wp'(\tilde{u})}{4\wp(\tilde{u})}+\frac{2}{\wp'(\tilde{u})}\right),
\end{align}
and accordingly,
\begin{align}
x_{\mathrm{HMM}}(\tilde{u})-2w_{\mathrm{HMM}}(\tilde{u})&=\frac{\tilde{\omega}}{8}\frac{\wp'(\tilde{u})}{\wp(\tilde{u})},\\
x_{\mathrm{HMM}}(\tilde{u})+8w_{\mathrm{HMM}}(\tilde{u})&=\frac{\tilde{\omega}}{2}\frac{\wp^2(\tilde{u})+1}{\wp'(\tilde{u})}.
\end{align}
On the other hand, we find from (\ref{Octa4 x}) and (\ref{Octa4 w}) that
\begin{align}
x(u)-w(u)&=-\frac{\omega}{2}\sqrt{\wp(u)},\\
x(u)+4w(u)&=\frac{\omega}{4}\frac{\wp'(u)}{\wp(u)}.
\end{align}
Thus, the equivalence we have to show turns out to be
\begin{align}
\frac{\tilde{\omega}}{8}\frac{\wp'(\tilde{u})}{\wp(\tilde{u})}&=-\frac{\omega}{2}\sqrt{\wp(u)},\label{equiv 1}
\end{align}
and
\begin{align}
\frac{\tilde{\omega}}{2}\frac{\wp^2(\tilde{u})+1}{\wp'(\tilde{u})}&=\frac{\omega}{4}\frac{\wp'(u)}{\wp(u)}.\label{equiv 2}
\end{align}

Due to the fact that $\displaystyle u=\omega(s+1)=\tilde{\omega}(s+1)/(1+i)=\tilde{\omega}(s+1)(1-i)/2=(1-i)\tilde{u}$, we notice from the addition theorem to the Weierstrass functions \cite{WhittakerWatson}
\begin{align}
\wp(u)=\wp(\tilde{u}-i\tilde{u})=\frac{1}{4}\left(\frac{\wp'(\tilde{u})-\wp'(-i\tilde{u})}{\wp(\tilde{u})-\wp(-i\tilde{u})}\right)^2-\wp(\tilde{u})-\wp(-i\tilde{u}).\label{addition thm}
\end{align} 
We also find  
\begin{align}
\wp(-i\tilde{u})&=-\wp(\tilde{u}),\label{lemma 1.1}\\
\wp'(-i\tilde{u})&=-i\wp'(\tilde{u}),\label{lemma 1.2}
\end{align}
from the definition of the Weierstrass function
\begin{align}
\wp(u)=\frac{1}{u^2}+\sum_{(m,n)\neq(0,0)}\left(\frac{1}{(u-\Omega_{m,n})^2}-\frac{1}{\Omega_{m,n}^2}\right),
\end{align}
where $\Omega_{m,n}=2m\omega_1+2n\omega_3$ and the fundamental half-periods for this case are $\omega_1=\omega$ and $\omega_3=i\omega$.
Thus, we observe from (\ref{addition thm}), (\ref{lemma 1.1}) and (\ref{lemma 1.2}) that
\begin{align}
\wp(u)=\frac{1}{4}\left(\frac{(1+i)\wp'(\tilde{u})}{2\wp(\tilde{u})}\right)^2
=\frac{i}{8}\left(\frac{\wp'(\tilde{u})}{\wp(\tilde{u})}\right)^2.\label{lemma 1.3}
\end{align}
A straightforward calculation then shows the equivalence (\ref{equiv 1}) with an appropriate choice of branch.
Next, we differentiate both sides of (\ref{lemma 1.3}) with respect to $u$, then find that
\begin{align}
\wp'(u)=\frac{i}{4}(1+i)\frac{\wp^2(\tilde{u})+1}{\wp^2(\tilde{u})}\wp'(\tilde{u}).\label{lemma 2}
\end{align}
The equivalence (\ref{equiv 2}) is a direct consequence of (\ref{lemma 2}).
Finally, the equivalence of the residue structure of the Nahm data (\ref{Octa4 bulk data}) and (\ref{Octa4 HMM data}) are confirmed at $s=\pm1$, which are necessary if we consider the monopole limits.

We therefore conclude that the octahedrally symmetric 4-caloron \cite{Ward2004} is a special case 
of the 4-caloron considered in section \ref{One parameter family}.

\section*{Acknowledgement}
The authors would like to thank Nobuyuki Sawado for his helpful comments and fruitful discussions.



\newpage

\begin{thebibliography}{0}
\bibitem{MantonSutcliffe}
 N.~Manton and P.~Sutcliffe,
  ``Topological Solitons'',
  Cambridge, UK: Cambridge University Press (2004). 

\bibitem{ShifmanYung2009}
M.~Shifman, A.~Yung,
  ``Supersymmetric Solitons'',
  Cambridge, UK: Cambridge University Press (2009). 

\bibitem{Dunajski}
 M.~Dunajski,
  ``Solitons, Instantons and Twisters'',
Oxford, UK: Oxford University Press (2010). 

\bibitem{Weinberg2012}
E.~J.~Weinberg,
  ``Classical Solutions in Quantum Field Theory'',
  Cambridge, UK: Cambridge University Press (2012). 

\bibitem{Shnir2018}
Y.~M.~Shnir,
  ``Topological and Non-Topological Solitons in Scalar Field Theories'',
  Cambridge, UK: Cambridge University Press (2018). 






\bibitem{HS} B.J.~Harrington and H.~K.~Shepard, Phys. Rev. D \textbf{17} (1978) 2122;
 \textit{ibid.} \textbf{18} (1978) 2990.

\bibitem{PS} M.~K.~Prasad and C.~M. Sommerfield, Phys. Rev. Lett. \textbf{35} (1975) 760-762.

\bibitem{Bogo} E.~B.~Bogomol'nyi, Sov. J. Nucl. Phys. \textbf{24} (1976) 449-454.


\bibitem{Shnir2005}
Y.~Shnir,
  ``Magnetic Monopoles'',
Berlin Heidelberg, Germany: Springer-Verlag (2005). 



\bibitem{Skyrme} T.~H.~R.~Skyrme, Nucl. Phys.\textbf{31} (1962) 556.

\bibitem{AtiyahManton} M.~F.~Atiyah, N~.S~.Manton, Phys. Lett. \textbf{B222} (1989) 438.

\bibitem{Sutcliffe2010} P.~Sutcliffe, J. High Energy Phys.\textbf{08} (2010) 019.

\bibitem{Sutcliffe2015}
  P.~Sutcliffe,
  Mod.\ Phys.\ Lett.\ B {\bf 29} (2015) no.16,  1540051.


\bibitem{SakaiSugimoto} T.~Sakai, S.~Sigimoto, Prog. Theor. Phys. \textbf{113} (2005) 843.


\bibitem{Cork2018} J.~Cork,J. High Energy Phys. \textbf{11} (2018) 137.

\bibitem{LeeseManton1994} R.~A.~Leese, N.~S.~Manton,  Nucl. Phys. \textbf{A572} (1994) 575.

\bibitem{BattyeSutcliffe1997} R.~M.~Battye, P.~M.~Sutcliffe,  Phys. Rev. Lett. \textbf{79} (1997) 363;
  Nucl.\ Phys.\ B {\bf 510} (1998) 507.


\bibitem{BattyeSutcliffe2001} R.~M.~Battye, P.~M.~Sutcliffe,  Phys. Rev. Lett. \textbf{86} (2001) 3989.

\bibitem{BattyeSutcliffe2002} R.~M.~Battye, P.~M.~Sutcliffe,   Rev. Math. Phys. \textbf{14} (2002) 29.


\bibitem{BraatenTownsentCarson1990} E.~Braaten, S.~Townsent, L.~Carson,  Phys. Lett. \textbf{B235} (1990) 147.


\bibitem{FeistLauManton2013} D.~T.~J.~Feist, P.~H.~C.~Lau, N.~S.~Manton,  Phys. Rev. \textbf{D87} (2013) 085034.

\bibitem{HoughtonMantonSutcliffe1998}
  C.~J.~Houghton, N.~S.~Manton, P.~M.~Sutcliffe, Nucl. Phys. \textbf{B510} (1998) 507.


\bibitem{SingerSutcliffe1999} M.~A.~Singer, P.~M.~Sutcliffe,   Nonlinearity \textbf{12} (1999) 987.

\bibitem{Sutcliffe2004}  P.~M.~Sutcliffe,   Proc. Roy. Soc. Lond. \textbf{A460} (2004) 2903.

\bibitem{AllenSutcliffe2013} J.~P.~Allen, P.~M.~Sutcliffe,   J. High Energy Phys. \textbf{05} (2013) 63.


\bibitem{Ward2004}
R.~S.~Ward, Phys.\ Lett.\ {\bf B582} (2004) 203.




\bibitem{HMM1995}
N.~J.~Hitchin, N.~S.~Manton, M.~K.~Murray, Nonlinearlity {\bf 8} (1995) 661.

\bibitem{HoughtonSutcliffe1996CMP}
C.~J.~Houghton, P.~M.~Sutcliffe, Commun.\ Math.\ Phys. {\bf 180} (1996) 343.

\bibitem{HoughtonSutcliffe1996Nonl}
C.~J.~Houghton, P.~M.~Sutcliffe, Nonlinearlity {\bf 9} (1996) 385.





\bibitem{Nahm1980}
  W.~Nahm,
  Phys.\ Lett.\ B {\bf 90} (1980) 413; 
\textit{ibid.} {\bf 93} (1980) 42.

  
  

\bibitem{Hitchin1983} N.~J.~Hitchin, 
Comm. Math. Phys. \textbf{89} (1983) 145-190.

  
\bibitem{Nahm1984} W.~Nahm,  Springer Lecture Notes in Phys. \textbf{201}(1984) 189-200. 




\bibitem{BNvB}
  F.~Bruckmann, D.~Nogradi and P.~van Baal,
  Nucl.\ Phys.\ B {\bf 666} (2003) 197-229;
   \textit{ibid.} {\bf 698} (2004) 233-254.



\bibitem{Harland}
D.~Harland,   J.\ Math.\ Phys.\  {\bf 48} (2007) 082905.



\bibitem{Kato2018} T.~Kato, A.~Nakamula, K.~Takesue, J. High Energy Phys.\textbf{06} (2018) 024.




\bibitem{NakamulaSakaguchi}
A.~Nakamula, J.~Sakaguchi,
  J.\ Math.\ Phys.\  {\bf 51} (2010) 043503.


\bibitem{Cork2017} J.~Cork, J.~Math.~Phys.\textbf{59}(2018)062902.



\bibitem{MNST}
D.~Muranaka,\ A.~Nakamula,\ N.~Sawado,\ K.~Toda,
Phys.\ Lett.\ {\bf B703} (2011) 498.




\bibitem{NakamulaSawado}
A.~Nakamula, N.~Sawado,
Nucl. Phys. {\bf B868} (2013) 476.

\bibitem{NakamulaSawadoTakesue}
A.~Nakamula, N.~Sawado, K.~Takesue
J. Phys. Conf. Ser. {\bf 563} (2014) 012032.




\bibitem{ADHM} M.~F.~Atiyah, V.~G.~Drinfeld, N.~J.~Hitchin and Yu.I.~Manin, 
Phys. Lett. A \textbf{65} (1978) 185-187.





\bibitem{KvB}
T.~C.~Kraan, P.~van Baal, Nucl.\ Phys.\ {\bf B533} (1998) 627.

\bibitem{LL}
K.~Lee, C.~Lu, Phys.\ Rev.\ {\bf D58} (1998) 025011.






\bibitem{GPY} D.~Gross, R.~D.~Pisarski and L.~G.~Yaffe, Rev. Mod. Phys.\textbf{53} (1981) 43.

\bibitem{ShifmanUnsal} M.~Shifman and M.~\"Unsal, Nucl. Phys. B, Proc. Suppl \textbf{186} (2009) 235.

\bibitem{Diakonov} D.~Diakonov, Nucl. Phys. B, Proc. Suppl \textbf{195} (2009) 5.

\bibitem{DGPS} D.~Diakonov, N.~Gromov, V.~Petrov and S.~Slizovskiy, Phys. Rev. D \textbf{70} (2004) 036003.

\bibitem{PoppitzSchaferUnsal} E.~Poppitz, T.~Schafer and M.~\"Unsal, J. High Energy Phys. 1303 (2013) 087.




\bibitem{DiaGro} D.~Diakonov and N.~Gromov, Phys. Rev. D \textbf{72} (2005) 025003.

\bibitem{GroSli} N.~Gromov and S.~Slizovskiy, Phys. Rev. D \textbf{73} (2006) 025022.

\bibitem{Sli} S.~Slizovskiy, Phys. Rev. D \textbf{76} (2007) 085019.


\bibitem{DiakonovPetrov} D.~Diakonov and V.~Petrov, Phys. Rev. D \textbf{76} (2007) 056001.





\bibitem{GillardSutcliffe2010} M.~Gillard, P.~Sutcliffe, J. Math. Phys.. \textbf{51} (2010) 122305. 

\bibitem{Foster2012} D.~Foster, J. High. Energy. Phys.. \textbf{12} (2012) 081. 



\bibitem{BattyeHaberichter2013} R.~A.~Battye, M.~Haberichter, Phys. Rev. \textbf{D87} (2013) 105003. 


\bibitem{SamoilenkaShnir2018} A.~Samoilenka, Y.~Shnir, Phys. Rev. \textbf{D97} (2018) 105014. 



\bibitem{WhittakerWatson}
  E.~T.~Whittaker and G.~N.~Watson,
  ``A course of modern analysis (4th Ed.),''
  Cambridge, UK: Cambridge University Press (1927). 







\end{thebibliography}
\end{document}